\documentclass[nofootinbib,prd,preprintnumbers,groupedaddress]{revtex4}
\usepackage{graphicx}
\usepackage{amssymb}
\usepackage{amsmath}
\usepackage{color}
\usepackage{rotating}
\usepackage{subfig}
\usepackage{verbatim}
\usepackage[normalem]{ulem}
\usepackage[colorlinks=true,
            linkcolor = magenta,
            citecolor = green]{hyperref}

\begin{document}
\title{Lepton universality violation and lepton flavor conservation in $B$-meson decays}

\author{R.~Alonso$^{1}$}
\author{B.~Grinstein$^{1}$}
\author{J. Martin Camalich$^{1,2}$}

\affiliation{
$^1$Dept. Physics, University of California, San Diego, 9500 Gilman Drive, 
La Jolla, CA 92093-0319, USA\\
$^2$PRISMA Cluster of Excellence Institut f\"ur Kernphysik, 
Johannes Gutenberg-Universit\"at Mainz, 55128 Mainz, Germany}

\begin{abstract}
Anomalies in (semi)leptonic $B$-meson decays present interesting patterns
that might be revealing the shape of the new physics to come. In order to understand
the experimental data, we explore symmetry arguments that lead to the hypothesis of 
minimal flavor violation. In particular, under the assumption of negligible neutrino
mass effects in charged lepton processes, the presence of lepton universality violation
without lepton flavor violation naturally arises. This can account for a deficit of
$B^+\to K^+\mu\mu$ over $B^+\to K^+ee$ decays with new physics coupled predominantly to
muons and a new physics scale of a few TeV. A generic prediction of this scenario is the modification of processes involving the
third generation. In particular, accounting for the above ratio implies a large enhancement, by a factor $\sim10^3$ with respect to
the standard model, of all the $b\to s\tau\tau$ decay rates. Although these are still below
current experimental limits, they should be easily at reach in future experiments at $B$-factories.
Another important consequence is the prediction
of sizable effects in charge-current $B$ tauonic decays which could also explain the
enhancements that have been observed in the $B\to D^{(*)}\tau\bar \nu$ and
$B^-\to\tau^-\bar \nu$ decays. For the most part, the study is carried out in 
an effective field theory framework with an underlying $SU(2)_L\times U(1)_Y$ symmetry 
that emphasizes the model-independent correlations arising between low- and high-energy observables.
For example, a connection between $B$-decays and top physics
is pointed out. To complement the discussion, all possible (spin 0 and 1) leptoquark
models are matched to the low-energy field theory so that the effective analysis can be
used to survey these candidates for new physics. These models also serve as concrete 
examples where the hypotheses of this work can be implemented.

\end{abstract}
\maketitle

\section{Introduction}

Flavor processes have long been known to be an extraordinary indirect probe into new physics (NP) reaching
as high in energies as thousands of TeVs. In fact, in the absence of direct evidence of new particles,
flavor physics may well spearhead the discovery of whatever theory lays beyond the Standard Model (SM).
The main body of flavor data presents an overwhelming agreement with the SM, although a few anomalies
have started to surface in $B$-meson decays. Individually, none of these tensions with the SM is statistically
significant yet. However, they could be pointing to structural patterns of the presumed NP that raise interesting questions
and are worth exploring. 

Among these anomalies, the most striking is a hint of violation of lepton universality in 
$b\to s\ell\ell$ rare decays. This shows up in a ratio of the rates of $B\to K\mu\mu$ over $B\to Kee$,
called $R_K$, that is very accurately predicted to be 1 in the SM~\cite{Hiller:2003js,Bobeth:2007dw} and was
measured to be $R_K=0.745^{+0.090}_{-0.074}\pm0.036$ by the LHCb collaboration~\cite{Aaij:2014ora}. This represents a $2.6\sigma$
tension not only with the SM but also with lepton universality. Furthermore, different observables
produced by the same elementary quark transition, $b\to s\ell\ell$, also present deviations from the SM. 
These include analyses of the angular observables in the disintegration $B\to K^*\mu\mu$ at low $q^2$~\cite{Aaij:2013qta} 
and high $q^2$~\cite{Horgan:2013pva,Horgan:2013hoa} (where $q^2$ is the invariant dilepton mass squared) and in some branching
fractions~\cite{Khodjamirian:2012rm,Aaij:2013aln,Aaij:2013pta,Aaij:2014pli}. Finally,
tensions also appear in tauonic charge-current $B$ decays that could  be pointing to new 
lepton non-unitary charged current interactions. These involve $b\to c \tau\bar\nu$ and $b\to u \tau\bar\nu$
exclusive decay rates. In particular, in $B\to D^{(*)}\tau\bar \nu$ an excess with respect to the SM 
is observed with a $3.6\sigma$ significance~\cite{Randall:1993qg,Boyd:1995pq,Matyja:2007kt,Bozek:2010xy,Lees:2013uzd,Fajfer:2012vx,Becirevic:2012jf} 
while another could be manifesting in $B^-\to \tau^-\bar \nu$~\cite{Agashe:2014kda}. The appearance of all
the aforementioned anomalies has prompted intense theoretical activity
(see for example~\cite{Fajfer:2012vx,Becirevic:2012jf,Celis:2012dk,Descotes-Genon:2013wba,Beaujean:2013soa,Lyon:2014hpa,Alonso:2014csa,Hiller:2014yaa,Ghosh:2014awa,Hurth:2014vma,Altmannshofer:2014rta,
Glashow:2014iga,Jager:2014rwa,Bhattacharya:2014wla,Crivellin:2015mga,Sierra:2015fma,Boucenna:2015raa,Crivellin:2015era}).   
 
From a broad perspective, the immediate questions about these anomalies are: why would NP appear
in these processes and what other deviations from the SM can we expect if they turn out
to be real effects. In particular, the $R_K$ anomaly involves quark flavor violation (QFV)
and lepton universality violation (LUV)  in the same neutral-current process. In our view, this is a rather peculiar
situation for two reasons: $(i)$ Lepton flavor violation (LFV) is widely believed to be a much better
probe of new physics than LUV, yet we do not have evidence of the former; $(ii)$ the process involves
nonstandard QFV and LUV none of which have been detected separately.

Regarding the first question, one expects, on general grounds, that any new source of flavor beyond the SM would 
produce both universality and flavor violation. In this case the new interaction responsible for the LUV signal
in $R_K$ should produce LFV in $b\to s\ell\ell'$ transitions~\cite{Glashow:2014iga}.  
However, as we intend to show in Sec.~\ref{SEC2}, symmetry considerations can lead to a situation where lepton
universality is violated but lepton flavor is conserved.

The second question above can be addressed with some new quark-lepton interaction. 
Our study will be, to a large extent, developed within the framework of Effective Field Theories (EFT),
comprising Sec.~\ref{SEC3}. We will also study a specific class of leptoquark models,
which will serve as concrete examples, in Sec.~\ref{Sec:lepqtoquarkillustrated} and the appendix~\ref{AppLQ}.
Note that the effects of the leptoquark models in flavor observables or $R_K$ have been addressed earlier in the literature~\cite{Buchmuller:1986zs,Davidson:1993qk,Hewett:1997ce,Sakaki:2013bfa,Hiller:2014yaa,Gripaios:2014tna,Sahoo:2015wya,Varzielas:2015iva,Becirevic:2015asa}. 
The EFT formalism will be introduced in Sec.~\ref{SEC1} where we will emphasize the role of 
the unbroken $SU(2)_L\times U(1)_Y$ symmetry to construct the most general NP operators and to derive model-independent relations
between different low- and high-energy observables. The experimental data relevant for our discussion
is reviewed in Sec.~\ref{Sec:Experiment}.

\section{The high- and low-energy effective theories}
\label{SEC1}
\subsection{The low-energy effective Lagrangians}

Flavor-changing neutral currents are induced at the quantum level and
are GIM~\cite{Glashow:1970gm} suppressed in the SM. 
In particular, $\Delta B=1$ decays are described by the effective Lagrangian~\cite{Grinstein:1988me,Buchalla:1995vs,Chetyrkin:1996vx}:
\begin{equation}
\label{eq:Leffnc}
  {\cal{L}}_{\rm n.c.}=-
  \frac{4 G_F}{\sqrt{2}}\sum_{p=u,c}\lambda_{pi}\left(C_1 \mathcal{O}_1^p+C_2\mathcal{O}_2^p +C_\nu \mathcal{O}_\nu+\sum_{k=3}^{10} C_k  \mathcal{O}_k\right),
\end{equation}
where Fermi's constant is, in terms of the electroweak vev, $G_F=1/(\sqrt2 v^2)$, $v=246$GeV, the chiral projectors are defined as usual, $P_{R,L}=(1\pm\gamma_5)/2$,  $\lambda_{pi}=V_{pb} V_{p i}^\ast$ with $i$ running through $s$ and $d$
quarks, and where the $C_{1\ldots 10}$ are the Wilson coefficients of the effective theory.
The $\mathcal{O}^p_{1,2}$ and $\mathcal{O}_{3-6}$ are the ``current-current'' and ``QCD penguin''
four-quark operators;  $\mathcal{O}_7$ and $\mathcal{O}_8$ encapsulate the effects of the 
``electromagnetic'' and ``chromo-magnetic'' penguins~\cite{Buchalla:1995vs}. Finally, $\mathcal{O}_9$, $\mathcal{O}_{10}$
and $\mathcal{O}_\nu$ are semi-leptonic operators involving either charged leptons or neutrinos and
will be the relevant ones for our study.
These are defined as:
\begin{align}
\mathcal{O}_9  &= 
    \frac{e^2}{(4 \pi)^2} [\bar d_i \gamma_\mu  P_ Lb][\bar{l} \gamma^\mu l],&
  \mathcal{O}_{10}  = &
    \frac{e^2}{(4 \pi)^2} [\bar d_i \gamma_\mu  P_L b][\bar{l} \gamma^\mu \gamma_5 l], &
    \mathcal{O}_\nu =&\frac{e^2}{(4 \pi)^2} [\bar d_i \gamma^\mu P_L b] [\bar \nu\gamma^\mu(1-\gamma_5) \nu],
\label{eq:semilOps}
\end{align}
where $b$ is the bottom quark field, $d_i$ stands for the strange and down quarks, $d_i=s,d$, and  $l,\nu$  are the charged lepton and neutrino,
respectively. Chirally-flipped ($b_{L(R)}\rightarrow b_{R(L)}$) versions
of all these operators are negligible in the SM, although they need not be so in
NP scenarios. In addition, NP can generate scalar and tensor
operators~\cite{Bobeth:2007dw},
\begin{align}\label{eq:nonSM:scalars} 
  \mathcal{O}_S^{(\prime)} & = \frac{e^2}{(4 \pi)^2} [\bar d_i P_{R(L)} b] [\bar{l} l], &
  \mathcal{O}_P^{(\prime)}  &= \frac{e^2}{(4 \pi)^2} [\bar d_i P_{R(L)} b] [\bar{l} \gamma_5 l],
\\  \label{eq:nonSM:tensors}
  \mathcal{O}_T   & = \frac{e^2}{(4 \pi)^2} [\bar d_i \sigma_{\mu\nu} b][\bar{l} \sigma^{\mu\nu} l], &
  \mathcal{O}_{T5} & = \frac{e^2}{(4 \pi)^2} [\bar d_i \sigma_{\mu\nu} b][\bar{l} \sigma^{\mu\nu} \gamma_5 l],
\end{align}
where  $\sigma^{\mu \nu} =  i\,[\gamma^\mu, \gamma^\nu]/2$.
The flavor index for leptons has been omitted, but we bear in mind that there is
an operator for every lepton flavor choice.

The charged current Lagrangian will also be necessary for our study. To leading order in $G_F$, the 
most general elementary charged-current Lagrangian mediating semileptonic decays reads~\cite{Cirigliano:2009wk,Cirigliano:2012ab}:
\begin{eqnarray}\label{eq:Leffcc}
{\cal L}_{\rm c.c} 
&=&
- \frac{4 G_F}{\sqrt{2}} \,V_{ib}\Big[(1+\epsilon_{L}^{ib})\,(\bar u_i\gamma^\mu P_L b)(\bar l\gamma_\mu  U P_L\nu)+
\epsilon_R^{ib}\,(\bar u_i\gamma^\mu P_R b)(\bar l\gamma_\mu U P_L\nu)\nonumber\\
&+&\epsilon^{ib}_{s_L}\,(\bar u_i P_L b)(\bar l\,U P_L\nu)
+\epsilon_{s_R}^{ib}\,(\bar u_i P_R b)(\bar l\, UP_L\nu)+\epsilon_T^{ib}(\bar u_i\sigma^{\mu\nu}P_L b)(\bar l\sigma_{\mu\nu}UP_L\nu)\Big]+{\rm h.c.}
\end{eqnarray}
where
$V$ is the CKM matrix, $u_i$ runs through $u$, $c$, and $t$ quarks, $U$ stands for the PMNS
matrix, lepton indices have not been made explicit for briefness and the Wilson coefficients $\epsilon$ quantify
deviations from the SM. The Lagrangian in eq.~(\ref{eq:Leffcc}) together with that in eq.~(\ref{eq:Leffnc}) with the addition
of the operators in eqs.~(\ref{eq:nonSM:scalars}-\ref{eq:nonSM:tensors})
constitute the most general low energy Lagrangian that describes
$B$-meson (semi-)leptonic decays with left-handed neutrinos.
\footnote{Note that similar operators with right-handed neutrinos do not interfere with the 
SM in the total decay rate (summed over final lepton polarizations). Therefore, in this case,
the dependence on the corresponding NP Wilson coefficients is quadratic instead of linear~\cite{Cirigliano:2012ab}.}

\subsection{The SM effective field theory}

If the relevant mass scale of NP, $\Lambda$, is larger than the electroweak vev,
we can integrate out the new particles in the unbroken phase and obtain operators
explicitly invariant under the SM gauge group: $SU(3)_c\times SU(2)_L\times U(1)_Y$. 
The effective field theory built with the most general set of operators will be referred to as 
the Effective Field Theory of the Standard Model (SMEFT) and relies on the expansion on the
ratio of the weak scale $v$ over the high energy scale $\Lambda$. The first terms in this expansion
are dimension five~\cite{Weinberg:1979sa} and dimension six operators~\cite{Buchmuller:1985jz,Grzadkowski:2010es}.
A particular advantage of the SMEFT is that it allows to treat a wide variety of phenomena spanning
different energy regimes, from  Higgs physics to kaon decays, in a systematic and model-independent fashion. In
the following, we assume that the electro-weak symmetry breaking is linearly realized, meaning that the Higgs doublet is 
treated as an elementary set of scalar fields. 
The non-linear realization would imply a larger set of operators at leading order~\cite{Buchalla:2012qq},
breaking the $SU(2)_L\times U(1)_Y$ relations of~\cite{Alonso:2014csa}.

 The  contributions that preserve lepton number are, at leading order, operators of dimension six,
$\mathcal{L}_{\rm NP}=\frac1{\Lambda^2}\sum_i C_i Q_i$, and the operators contributing
to (semi-)leptonic processes at low energies are of the Higgs-current times fermion-current
or four-fermion type~\cite{Grzadkowski:2010es}. Those containing a Higgs current nonetheless
induce{, at the $B$-meson scale and for neutral-current decays
either QFV or LUV but not both
in the same operator at leading order, so we will neglect them here. The four-fermion operators inducing $B$-meson (semi-)leptonic
rare decays are:
\begin{align}
Q_{\ell q}^{(1)}          =      & (\overline q \gamma^\mu q_L)(\bar \ell \gamma_\mu \ell_L) &
Q_{\ell q}^{(3)}          =      & (\overline q\vec{\tau} \gamma^\mu q_L)\cdot(\bar \ell \vec{\tau} \gamma_\mu \ell_L)\nonumber\\
Q_{\ell d}        =       & (\bar d \gamma^\mu d_R)(\bar \ell \gamma_\mu \ell_L)&
Q_{qe}     =          & (\overline q \gamma_\mu q_L)(\overline e \gamma^\mu e_R)   \nonumber\\
Q_{ed}   =                   & (\bar d_R \gamma^\mu d_R)(\overline e \gamma_\mu e_R) &
Q_{\ell edq} =& (\bar \ell_L e_R)(\overline d_R q){\rm+h.c.} \label{eq:Leffdown}
\end{align}
where color and weak-isospin indices are omitted, $\tau^I$ stand for the Pauli matrices in
$SU(2)_L$-space, $q$ and $\ell$ are the quark and lepton doublets respectively, $q=(u_L,d_L)$ and
$\ell=(\nu_L,l_L)$ and $e_R$ and $d_R$ are the right-handed charged leptons and down-type quarks. 
Contributions to charged-current or up-quark flavor-neutral decays can also be generated by:
\begin{align}
Q_{lequ}^{(1)} =& (\bar \ell e_R) (\overline q_L u_R) +{\rm h.c.} & Q_{lequ}^{(3)} =& (\bar \ell \sigma_{\mu\nu} e_R)(\overline q_L \sigma^{\mu\nu}u_R)+{\rm h.c.}\nonumber\\
Q_{eu} =& (\overline e \gamma_\mu e_R)(\bar u \gamma^\mu u_R) &
Q_{\ell u}=& (\bar \ell \gamma_\mu \ell)(\bar u \gamma^\mu u_R) \label{eq:Leffup}
\end{align}
where $u_R$ stands for up-type right-handed quarks and flavor indices have also been omitted for brevity. In general we will use \emph{greek letters} for lepton flavor
indices and \emph{latin letters} for quark flavor indices, using the notation: 
$(Q_{\ell q}^{(1)})^{\alpha\beta,ij}=(\overline q^j \gamma^\mu q_L^i)(\bar \ell^\alpha \gamma_\mu \ell^\beta)$.
Each operator appears in the Lagrangian with a coefficient $C_i/\Lambda^2$, such that at low energy one cannot tell the scale $\Lambda$ from the dimensionless
coefficient $C_i$. However we will not consider arbitrary values of the two parameters for a fixed value of the ratio; we will consider only perturbative coefficients,
in particular $C\leq 4\pi$. An illustration of the implication of this limit on the couplings of specific models will be presented in Sec.\ref{Sec:lepqtoquarkillustrated}.

In order to connect to the effective Lagrangians at low energies, 
we transform from the interaction basis to the mass basis. In our convention 
this implies $q_{L,i}\to((V^\dagger\,u_L)_i,d_{L,i})$, $\ell_\alpha \to ((U\nu)_\alpha, l_\alpha)$ and
the right-handed fermions need not be rotated. Please note that this choice does not imply a loss of generality. 

The connection between the Lagrangians in eq.~(\ref{eq:Leffnc}) and that built with the operators
of eq.~(\ref{eq:Leffdown}) is (for complete expressions see~\cite{Alonso:2014csa,Buras:2014fpa}),
\begin{align}
\delta C_9&=\frac{4\pi^2}{e^2\lambda_{ti}}\frac{v^2}{\Lambda^2}\left(C_{qe}+C_{\ell q}^{(1)}+C_{\ell q}^{(3)} \right),&
\delta C_{10}&=\frac{4\pi^2}{e^2\lambda_{ti}}\frac{v^2}{\Lambda^2}\left(C_{qe}-C_{\ell q}^{(1)}-C_{\ell q}^{(3)} \right),\nonumber\\
\delta C_9'&=\frac{4\pi^2}{e^2\lambda_{ti}}\frac{v^2}{\Lambda^2}\left(C_{ed}+C_{\ell d}\right),&
\delta C_{10}'&=\frac{4\pi^2}{e^2\lambda_{ti}}\frac{v^2}{\Lambda^2}\left(C_{ed}-C_{\ell d}\right), \nonumber\\ \nonumber
\delta C_\nu&=\frac{4\pi^2}{e^2\lambda_{ti}}\frac{v^2}{\Lambda^2}\left(C_{\ell q}^{(1)}-C_{\ell q}^{(3)}\right),&
\delta C_\nu^\prime&=\frac{4\pi^2}{e^2\lambda_{ti}}\frac{v^2}{\Lambda^2}C_{\ell d},\\
\delta C_S&=-\delta C_P=\frac{4\pi^2}{e^2\lambda_{ti}}\frac{v^2}{\Lambda^2}C_{\ell e dq}, & \delta C_S^\prime&=\delta C_P^\prime=\frac{4\pi^2}{e^2\lambda_{ti}}\frac{v^2}{\Lambda^2}C_{\ell e dq}^\prime,
\label{eq:SMEFTnc}
\end{align}
where $C_{\ell e dq}^\prime$ corresponds to the hermitian of the operator $Q_{\ell edq}$ for the flavor 
entry $ji=bs$.
Note that as discussed in Ref.~\cite{Alonso:2014csa}, not all operators in eqs.~(\ref{eq:semilOps}-\ref{eq:nonSM:tensors}) are generated
or independent; in our particular case only 6 of the 10 operators are independent.
The  operator $Q_{\ell q}^{(3)}$ also contributes to $\mathcal{L}_{\rm c.c.}$,
\begin{equation}
\epsilon^{ij}_L=-\frac{v^2}{\Lambda^2}\sum_k\frac{V_{ik}}{V_{ij}}(C_{\ell q}^{(3)})_{kj},\label{eq:SMEFTcc}
\end{equation}
where we have omitted lepton-flavor indices. Note that contributions to $\epsilon^{ij}_R$ up to $\mathcal{O}(v^2/\Lambda^2)$
 can only
be generated by one of the Higgs-current operators, $i\tilde H ^\dagger D_\mu H\, \bar u\gamma^ \mu d_R$,
after integrating out the $W$ boson and, therefore, it respects lepton universality~\cite{Cirigliano:2009wk}. Contributions to
left-handed charged quark currents coupled to anomalous lepton charged currents via the exchange of a $W$ boson
have a negligible effect in meson decays due to the experimental constraints on the relevant $Wl \nu$ couplings that can be
derived from the weak boson decays~\cite{Antusch:2006vwa,Agashe:2014kda}.~\footnote{There is a notable exception in the $W\ell\nu$
couplings as LEP data contains a few-percent excess, at $\sim 2.5\sigma$, of tauonic decays
with respect to electronic or muonic. However, this is difficult to understand in the light of the per-mille-level
lepton-universality tests done with the purely leptonic $\tau$ decays (see~\cite{Filipuzzi:2012mg} for a
comprehensive analysis).} A corollary of this is that not only
for the neutral-current but also for the charged-current $B$ decays, any NP effect violating lepton universality
at $\mathcal{O}(v^2/\Lambda^2)$ must originate from the four-fermion operators of the SMEFT Lagrangian.

All the expressions included in this section describe the tree-level matching between 
the low- and high-energy EFT. The full analysis would imply running 
the coefficients of the operators in eq.~(\ref{eq:Leffdown}) from the high scale
$\Lambda$ to the electro-weak vev  (the full anomalous dimension matrix is given
in~\cite{Jenkins:2013zja,Jenkins:2013wua,Alonso:2013hga,Alonso:2014zka}), and then
down to the $B$-meson scale~\cite{Grinstein:1988me,Buchalla:1995vs,Chetyrkin:1996vx,Cirigliano:2009wk}. 
 
\section{Lepton Universality Violation without Flavor Violation}
\label{SEC2}

\label{Sec:MFV}
Symmetry considerations offer insight and robust arguments in particle physics. They explain
the absence of certain effects or their suppression with respect to others.
It seems therefore a good idea to pose the question of LUV and LFV in terms of symmetries: \emph{Is there
any symmetry that allows lepton universality violation  but conserves lepton flavor?} 
Yes, lepton family number: $U(1)_\tau\times U(1)_\mu\times U(1)_e$. This symmetry conserves tau, muon and electron number, although it allows for their respective couplings to differ from one another. Since this is the central point of this discussion, let us be more precise about the definition of the symmetry.

The gauge interactions of the SM respect a global flavor symmetry which in the case of leptons is $U(3)_\ell\times U(3)_{e}$. The symmetry transformation is a unitary rotation in generation-space
for each SM lepton having different quantum numbers, explicitly:
\begin{align}\label{LFS}
&SU(3)_\ell \times SU(3)_e\times U(1)_L\times U(1)_{e-\ell}\,, &
\ell_L&\sim(3,1)_{1,-1}\,, &
e_R&\sim(1,3)_{1,1}\,.
\end{align}
where we have grouped the global $U(1)$ symmetries into a vector rotation $U(1)_L$, which is the customary 
Lepton Number, and an axial rotation $U(1)_{e-\ell}$. The Yukawa interactions break this symmetry, leaving aside for a moment neutrino masses, they read:
\begin{align}\label{LY}
-\mathcal{L}_Y=\varepsilon_d \, \bar q_L \hat Y_d d_R H+\varepsilon_u\,\bar q_L \hat Y_u u_R \tilde H+\varepsilon_e\,\bar \ell_L \hat Y_e e_R H+{\rm h.c.}
\end{align}
where we have separated the Yukawa couplings into an overall flavor blind complex parameter, $\varepsilon_\psi$, and a normalized matrix that determines
the flavor structure along the lines of Ref.~\cite{Alonso:2011jd}. The relation to the usual notation
is
\begin{align}
Y_{e}&\equiv\varepsilon_e \hat Y_e\,, & \mbox{Tr}\left(\hat Y_e\hat Y_e^\dagger \right)&=1\,.\label{hatYe}
\end{align}
In particular this normalization sets $|\varepsilon_e|^2= y_e^2+y_\mu^2+y_\tau^2$.

For leptons, at this level, the presence of a Yukawa term breaks the symmetry although not completely. Indeed one can
use a unitary transformation in flavor space, which does not affect the rest of the Lagrangian, to make $Y_e$ diagonal: 
\begin{align}\label{LFTrns}
\ell_L&\to e^{i\theta_\ell}\hat U_\ell \ell_L\,,\quad e_R\to e^{i\theta_e}\hat U_e e_R\,;
&\varepsilon_e\to e^{i(\theta_e-\theta_\ell)}\varepsilon_e&=|\varepsilon_e|,
& \hat Y_e&\to\hat U_\ell^\dagger\hat Y_e\hat U_e=\frac{\sqrt{2}}{v|\varepsilon_e|}\mbox{diag}(m_e\,,m_\mu\,,m_\tau)\,, 
\end{align}
where $\hat U_{\ell,e}$ stand for special-unitary matrices, $\theta_{e,\ell}$ are global phases, and note that $\varepsilon_e$ only transforms under $U(1)_{\ell-e}$.
In this basis, is easy to see that there is an unbroken flavor symmetry:
\begin{align}
U(3)_\ell \times U(3)_e\, \to \,U(1)_\tau \times U(1)_\mu \times U(1)_e\,.
\end{align}
This is the definition of the symmetry referred to at the beginning of the section, and requires the introduction
of the mass basis for charged leptons as discussed above; it has been, indeed, long ago identified in the SM. 
Any other source of lepton-flavor symmetry breaking beyond the SM 
will be, in general, non-diagonal in the charged-lepton mass basis. In this case, we have the breakdown
of the leptonic flavor symmetry down to (possibly) $U(1)_L$ and both, LUV and LFV, would ensue~\cite{Glashow:2014iga}.  
On the other hand, if the NP \emph{explicitly} respects the $U(1)_\tau \times U(1)_\mu \times U(1)_e\,$ symmetry
to an approximate degree, then there can be universality violation but flavor transitions between
different generations are suppressed \emph{for charged leptons}.

This assumption has to be nonetheless confronted with two potential problems: \emph{i)} the fact that
neutrinos are massive making the symmetry not exact and, \emph{ii)} at a more theoretical level, why
would the NP flavor structure align with the charged lepton mass basis.

The presence of neutrino masses breaks the symmetry since angles in the mixing matrix connect different generations;
in other words, the conservation of this symmetry would require the charged lepton and neutrino masses to be simultaneously diagonalizable and hence a trivial mixing matrix. 
Our assumption requires that this source of breaking be negligible in the observables of interest and these involve
charged leptons. 
This seems most natural by looking at values of charged lepton vs neutrino masses ($m_e/m_{\nu}\gtrsim10^6$) and it also follows in specific models, {\it e.g.}, the generic type I seesaw, although there are particular models for which the hypothesis does not hold~\cite{Gavela:2009cd}. It also follows that we do not have to specify the mechanism
for neutrino mass generation, avoiding the ambiguities it entails.

As for the second point, even if there is nothing a priori unsustainable about assuming that the NP aligns with $\hat Y_e$, there is a simple
and furthermore predictive explanation: the source of flavor in the NP \emph{is} the charged lepton 
Yukawa coupling. 
The assumption that $\hat Y_e$ and $\varepsilon_e$ control the flavor structure of new physics follows
in Minimal Flavor Violation (MFV)~\cite{Chivukula:1987py,DAmbrosio:2002ex,Cirigliano:2005ck,Davidson:2006bd,Alonso:2011jd}.
The implementation of MFV is simply demanding the formal restoration of the flavor symmetry %
 treating the Yukawas as spurions. This procedure assigns transformation properties to Yukawa couplings
 so as to preserve the symmetry in the Yukawa interactions of eq.~(\ref{LY}); in particular for leptons and with the definitions of eq.~(\ref{LFTrns}), we have $\hat Y_e\sim(3,\bar 3)_{0,0}$ and $\varepsilon\sim(1,1)_{0,-2}$ where the lepton symmetry is that of eq.~(\ref{LFS}).
Similarly, preserving the lepton flavor symmetry in the Lagrangian built with the operators of eq.~(\ref{eq:Leffdown}) requires Yukawa insertions as follows:
\begin{align}
C_{\ell q}^{(1)}=&C_q^{(1)} \hat Y_e\hat Y_e^\dagger+\mathcal{O}((\hat Y_e\hat Y_e^\dagger)^2)  &C_{\ell q}^{(3)}=&C^{(3)}_q \hat Y_e\hat Y_e^\dagger+\mathcal{O}((\hat Y_e\hat Y_e^\dagger)^2) &
C_{\ell d}=&C_d \hat Y_e \hat Y_e^\dagger+\mathcal{O}((\hat Y_e\hat Y_e^\dagger)^2) \\ C_{qe }=&C_{q} \hat Y_e^\dagger \hat Y_e+\mathcal{O}((\hat Y_e\hat Y_e^\dagger)^2) &
C_{ed}=&C^\prime_d \hat Y_e^\dagger \hat Y_e+\mathcal{O}((\hat Y_e\hat Y_e^\dagger)^2) & C_{\ell e d q}=&C_{dq}\varepsilon_e \hat Y_e+\mathcal{O}(\varepsilon_e\hat Y_e^3\,,\varepsilon_e^3\hat Y_e) \label{CEFTMLFV}
\end{align}
where we have assumed a perturbative expansion in Yukawas, omitting the zeroth term since we are focusing on flavor effects. 
 We will consider the general case in the following sections. 
It is also worth remarking that only the operator $Q_{\ell e dq}$ is affected by an axial $U(1)_{e-\ell}$ phase rotation and therefore requires one power of $\varepsilon_e$. If we, in addition, assume MFV in the quark sector, the number of operators that induce QFV
reduces and the predictivity in quark flavor space increases:
\begin{align}
C_{\ell q}^{(1)}=&c_{\ell q}^{(1)}\,\hat Y_u\hat Y_u^\dagger\otimes \hat Y_e\hat Y_e^\dagger,  &C_{\ell q}^{(3)}=&c_{\ell q}^{(3)}\,\hat Y_u\hat Y_u^\dagger \otimes \hat Y_e\hat Y_e^\dagger, 
\\
C_{q e}=&c_{qe}\,\hat Y_u\hat Y_u^\dagger  \otimes \hat Y_e^\dagger \hat Y_e,
& C_{\ell e d q}=&c_{\ell eqd}\,\varepsilon_e\varepsilon_d^*\,\hat Y_d^\dagger\hat Y_u\hat Y_u^\dagger \otimes \hat Y_e. \label{CEFTMFV}
\end{align}
where with our normalization $|\varepsilon_d|^2=y_d^2+y_s^2+y_b^2$ and $|\varepsilon_u|^2=y_u^2+y_c^2+y_t^2$. 
Note that the symmetry argument dictating insertions of $\varepsilon_\psi$ naturally suppresses scalar operators
with respect to the current-current type of 4 fermion operators. 
On the other hand note that the operator's $Q_{ed}, Q_{\ell d}$ contributions to $b \to s$ transitions, whose quark-flavor coefficients would 
be $\hat Y_d^\dagger \hat Y_u\hat Y_u^\dagger \hat Y_d$,  are suppressed with respect to operators with left-handed quark currents by a factor $m_s/m_b$.
Finally we shall also note that the operators $Q_{\ell q}$ do induce
neutrino flavor violation, this however is much less constrained than charged lepton flavor violation,
specially for a four fermion operator that involves the $b$ quark.

\section{Experimental data}
\label{Sec:Experiment}
We describe in this section the experimental data that is useful for the discussion of the
scenarios with LUV in the MFV benchmarks described above.

\subsection{Rare exclusive $B_{d,s}$ (semi-)leptonic decays}

\subsubsection{The $R_K$ anomaly}

The LHCb measured the following lepton-universality ratio of the $B^+\rightarrow K^+ \ell\ell$ decay
in the bin $q^2\in[1,\,6]\,\text{GeV}^2$,
\begin{equation}
R_K\equiv \frac{\mathcal{B}\left(B^+\rightarrow K^+ \mu\mu \right)}{\mathcal{B}\left(B^+\rightarrow K^+ ee\right)}=0.745^{+0.090}_{-0.074}{\rm (stat)}\pm0.036 {\rm (syst)}.
\label{eq:RKexpt}
\end{equation}
The hadronic matrix elements cancel almost exactly in this ratio and $R_K$ is predicted to be approximately equal
to 1 in the SM~\cite{Bobeth:2007dw}. Therefore, a confirmation of this observation, which currently poses a $2.6\sigma$ discrepancy with the SM, would imply a clear manifestation of NP and LUV. Different
theoretical analyses show that this effect must be contained in the semileptonic operators 
$\mathcal{O}_{9,10}^{(\prime)}$ of the low-energy Lagrangian~\cite{Alonso:2014csa,Hiller:2014yaa,Ghosh:2014awa,Hurth:2014vma,Altmannshofer:2014rta}.
In the context of the SMEFT, the (pseudo)scalar ones are ruled out by the branching fraction of $B_s\to\ell\ell$ (see below) while 
tensor operators of dimension 6 mediating down-type quark transitions are forbidden by the $SU(2)_L\times U(1)_Y$ symmetry~\cite{Alonso:2014csa}. 

In the absence of the (pseudo)scalar and tensor contributions and neglecting, for the sake of clarity,
$m_\ell^2/q^2$, $q^2/m_B^2$ and $m_K^2/m_B^2$, the differential decay rate of $B\to K\ell\ell$ is,
\begin{equation}
\frac{d\Gamma}{d q^2}=\frac{G_F^2\,\alpha_e^2|\lambda_{ts}|^2\,m_B^3}{1536\pi^5}\,f_+^2\left(|C_9+C_9^\prime+2\frac{\mathcal{T}_K}{f_+}|^2+|C_{10}+C_{10}^\prime|^2\right), \label{eq:BKtoll}
\end{equation}
where $f_+$ is a ($q^2$-dependent) hadronic form factor and $\mathcal{T}_K$ is a $q^2$-dependent
function accounting for the (lepton universal) contribution of a virtual photon to
the decay~\cite{Bobeth:2007dw,Beneke:2001at}. Taking into account that $C_9^{\rm SM}(m_b)=4.24\simeq-C_{10}^{\rm SM}$,
inspection of eq.~(\ref{eq:BKtoll}) shows that the $R_K$ anomaly requires any suitable combination
of the scenarios: 
\begin{align}
&\delta C_9^\mu-\delta C_9^e\in[-1,0],&\delta C_{10}^\mu-\delta C_{10}^e\in[0,1],\nonumber \\
&\delta C_9^{\mu\prime}-\delta C_9^{e\prime}\in[-1,0],&\delta C_{10}^{\mu\prime}-\delta C_{10}^{e\prime}\in[0,1].\label{eq:WCRK}
\end{align}

\subsubsection{Anomalies in the angular distribution of $B\to K^*\mu^+\mu^-$}

The $ B \to K^*(\to K\pi) \ell^+ \ell^-$ is a four body decay with a rich kinematic
structure that offers excellent opportunities to search for NP~(see e.g. \cite{Altmannshofer:2008dz,
Bobeth:2010wg,Egede:2010zc,DescotesGenon:2012zf,Jager:2012uw} and references therein). In fact,
a complete angular analysis of (1 fb$^{-1}$) data collected by the LHCb in the muonic channel 
showed a $3.7\sigma$ discrepancy with the SM in an angular observable called $P_5'$~\cite{Aaij:2013qta}.
Potential discrepancies have also been noted in other observables and different global analyses agree
that the tensions can be ascribed to a negative NP contribution to $C_9^{\mu}$~\cite{Descotes-Genon:2013wba,Altmannshofer:2013foa,
Beaujean:2013soa,Hurth:2013ssa,Altmannshofer:2014rta},
\begin{equation}
\delta C_9^\mu\simeq-1, 
\end{equation}
or within a (left-handed) scenario where ~\cite{Altmannshofer:2014rta},
\begin{equation}
\delta C_9^\mu=-\delta C_{10}^\mu\simeq-0.5,
\end{equation}
Note that these modifications are compatible with the possible scenarios to accommodate $R_K$ in
eq.~(\ref{eq:WCRK}) and also discard alternatives based on large values of the Wilson coefficients,
$C_{9,10}^{\rm SM}+\delta C_{9,10} =-C_{9,10}^{\rm SM}$.}
Indeed, complementarity of these NP interpretations with the measurements of $R_K$ and $B_s\to\mu\mu$
can be found in~\cite{Alonso:2014csa,Hiller:2014yaa,Ghosh:2014awa,Hurth:2014vma,Altmannshofer:2014rta}.
Interestingly, a recent angular analysis of the full 3 fb$^{-1}$ data set collected by the LHCb ratifies
the discrepancy with the SM~\cite{LHCb:2015dla,Altmannshofer:2015sma}. It is important to stress, though,
that it is not clear yet if the tensions can be accommodated in the SM by means of a not-fully-understood hadronic
effect (see for recent discussions~\cite{Jager:2012uw,Lyon:2014hpa,Descotes-Genon:2014uoa,Jager:2014rwa,Straub:2015ica}).

\subsubsection{Observation of $B_{d,s}\to\mu\mu$}

An important constraint on the $b\to s\mu\mu$ operators comes from the observation of $B_s\to\mu\mu$~\cite{CMS:2014xfa},
which has a branching fraction smaller but in good agreement (compatible at $1.2\sigma$) with the SM prediction~\cite{Bobeth:2013uxa}:
\begin{equation}
\overline{\mathcal{B}}_{s\mu}^{\rm expt}=2.8^{+0.7}_{-0.6}\times10^{-9},\hspace{1cm}\overline{\mathcal{B}}_{s\mu}^{\rm SM}= 3.65(23)\times10^{-9}.
\end{equation}
These modes are chirally suppressed and they induce strong bounds on the 
(pseudo)scalar operators~\cite{Alonso:2014csa}. There is a contribution from the operators $\mathcal{O}_{10}^{(\prime)}$ which reads
\begin{equation}
\overline{R}_{s\mu}=\frac{\overline{\mathcal{B}}_{s\mu}^{\rm expt}}{\overline{\mathcal{B}}_{s\mu}^{\rm SM}}=
\frac{1+\mathcal{A}^{\mu\mu}_{\Delta \Gamma}\,y_s}{1+y_s}\big|\frac{C_{10}^\mu-C_{10}^{\mu\prime}}{C_{10}^{\rm SM}}\big|^2, \label{eq:R}
\end{equation}
where $y_s=\tau_{B_s}\Delta\Gamma_s/2$, $\mathcal{A}^{\mu\mu}_{\Delta \Gamma}$ is the 
mass eigenstate rate asymmetry~\cite{DeBruyn:2012wk} and where we have explicitly indicated
the lepton-flavor dependence of the Wilson coefficients. Taking into account that $C_{10}^{\rm SM}=-4.31$, a contribution
as large as:
\begin{equation}
 \delta C_{10}^\mu-\delta C_{10}^{\mu\prime}\simeq0.5\gtrsim0\label{eq:WCBsll}
\end{equation}
improves the agreement with the measurement.
A similar constraint on the $b\to d\mu\mu$ operators stems from the observation, 
with a significance of $3.2\sigma$, of the $B_d\to\mu\mu$ decay~\cite{CMS:2014xfa}: 
\begin{equation}
\overline{\mathcal{B}}_{d\mu}^{\rm expt}=3.9^{+1.6}_{-1.4}\times10^{-10},\hspace{0.5cm}\overline{\mathcal{B}}_{d\mu}^{\rm SM}=1.06(9)\times10^{-10},
\end{equation}
which shows an excess of $2.2\sigma$ with respect to the SM prediction. 
Generalizing the formulae introduced above for $B_s\to\mu\mu$ and having already discarded (pseudo)scalar operators, this measurement allows for
contributions of the same order and sign as the SM one:
\begin{equation}
\delta C_{10}^\mu-\delta C_{10}^{\mu\prime}\simeq C_{10}^{\rm SM}<0,\label{eq:WCBdll}
\end{equation}
where the Wilson coefficient corresponds to a different quark-flavor transition as those in eq.~(\ref{eq:WCBsll}). However,
the two sets can be connected by flavor symmetries, like for instance through the ratio~\cite{CMS:2014xfa}: 
\begin{equation}
\mathcal{R}=\frac{\overline{\mathcal{B}}_{d\mu}^{\rm expt}}{\overline{\mathcal{B}}_{s\mu}^{\rm expt}}=0.14^{+0.08}_{-0.06},\label{BdBs}
\end{equation}
which is at $2.3\sigma$ above the SM \textit{and} the MFV prediction, 
$\mathcal{R}^{\rm MFV}=\mathcal{R}^{\rm SM}=0.0295^{+0.0028}_{-0.0025}$~\cite{Bobeth:2013uxa}. The MFV prediction follows
in particular if one uses MFV in the quark sector to accommodate the anomaly in $R_K$.

\subsubsection{Tauonic decays}

The rare $b\to s\tau\tau$ transitions are poorly constrained (see~\cite{Bobeth:2011st}
for a comprehensive analysis). We focus here on the current experimental limits in the
$B_s\to\tau\tau$ and $B\to K\tau\tau$ decays which give the best 
bounds on the underlying semileptonic operators~\cite{Bobeth:2011st}:
\begin{align}
&\overline{\mathcal{B}}_{s\tau}^{\rm SM}= 7.73\pm0.49\times10^{-7}~\text{\cite{Bobeth:2013uxa}},&
&\mathcal{B}_{s\tau}^{\rm expt}<3\%~\text{\cite{Bobeth:2011st}}\nonumber\\
&\mathcal{B}(B^+\to K^+\tau\tau)^{\rm SM}=1.44(15)\times10^{-7}~\text{\cite{Bouchard:2013mia}},&
&\mathcal{B}(B^+\to K^+\tau\tau)^{\rm expt}<3.3\times10^{-3}~\text{\cite{Flood:2010zz}}, \label{eq:btostautau}
\end{align}
where the experimental limits are at 90$\%$ C.L. As described in~\cite{Bobeth:2011st}, this
leads to constraints on $C_{9,10}^\tau$ not better than $C_{9,10}^\tau\lesssim2\times10^3$.

\subsubsection{Rare exclusive $b\to s\nu\bar\nu$ decays}

The exclusive decays into neutrinos have been searched for in the $B$-factories 
leading to stringent experimental limits (90$\%$ C.L.):
\begin{eqnarray}
&&\mathcal{B}(B^+\to K^+\nu\bar\nu)<1.7\times10^{-5}~\text{\cite{Lees:2013kla}},\nonumber\\
&&\mathcal{B}(B^0\to K^{*0}\nu\bar\nu)<5.5\times10^{-5}~\text{\cite{Lutz:2013ftz}},\nonumber\\
&&\mathcal{B}(B^+\to K^{*+}\nu\bar\nu)<4.0\times10^{-5}~\text{\cite{Lutz:2013ftz}},\label{eq:bsnunuexpt}
\end{eqnarray}
which are an order of magnitude larger than the SM predictions~\cite{Buras:2014fpa}. This is better 
expressed normalizing the decay rate with respect to the SM:
\begin{eqnarray}
R_{K^{(*)}\nu}=\frac{\mathcal{B}(B\to K^{(*)}\nu\bar\nu)}{\mathcal{B}(B\to K^{(*)}\nu\bar\nu)^{\rm SM}}, 
\end{eqnarray}
so that~(\ref{eq:bsnunuexpt}) implies~\cite{Buras:2014fpa}:
\begin{eqnarray}
R_{K\nu}<4.3,\hspace{1cm}R_{K^{*}\nu}<4.4,\label{eq:bsnunuRs} 
\end{eqnarray}
at 90\% C.L. These bounds are translated into constraints of the Wilson coefficients. 
For instance, assuming for simplicity that $C_\nu^\prime=0$ we have: 
\begin{eqnarray}
R_{K^{(*)}\nu}=\frac{|C_\nu|^2}{|C_\nu^{\rm SM}|^2},
\end{eqnarray}
where $C_\nu^{\rm SM}\simeq-6.35$. (For the slightly more involved expressions 
including $C_\nu^\prime$ see~\cite{Buras:2014fpa}).

\subsection{Semi-leptonic $B$-meson and top quark decays}

\subsubsection{The $B\to D^{(*)}\tau\nu$ anomalies}

If the spectrum of the decay $B\to D^{(*)}\mu\nu$ is measured, the decay $B\to D^{(*)}\tau\nu$
can be predicted with reduced theoretical input~\cite{Randall:1993qg,Boyd:1995pq,Fajfer:2012vx,Becirevic:2012jf}. In particular,
the ratio of the two decay rates,
\begin{eqnarray}
R_{D^{(*)}}=\frac{\mathcal{B}(\bar B\to D^{(*)}\tau\bar\nu_\tau)}{\mathcal{B}(\bar B\to D^{(*)}\mu\bar\nu_\mu)},
\end{eqnarray}
can be given accurately in the SM~\cite{Fajfer:2012vx,Becirevic:2012jf}:
\begin{eqnarray}
R_{D}^{\rm SM}=0.296(16),\hspace{1cm}R_{D^{*}}^{\rm SM}=0.252(3). 
\end{eqnarray}
Measurement of these modes have been reported by the BaBaR~\cite{Lees:2013uzd} and 
Belle~\cite{Matyja:2007kt,Bozek:2010xy} collaborations and an average of the experimental results
gives~\cite{Sakaki:2013bfa}:
\begin{equation}
 R_{D}^{\rm expt}=0.421(58),\hspace{1cm}R_{D^{*}}^{\rm expt}=0.337(25),
\end{equation}
which amounts to a combined 3.5$\sigma$ discrepancy with the SM. A possible explanation 
of this signal is a LUV contribution to the $V-A$ coupling:
\begin{equation}
\epsilon_L^{cb,\tau}-\epsilon_L^{cb,l}\sim 0.15,\hspace{1cm}l=e,\,\mu,\label{eq:vl23expt}
\end{equation}
although not by an equivalent LUV Wilson coefficient from $V+A$ quark currents, $\epsilon_R^{cb}$, as
these can only arise, at leading order, from Higgs-current type of operator in the SMEFT. Finally, it
is interesting to note that LUV is not required to explain the signal because contributions
from $\epsilon_{s_L,s_R}^{cb}$ or $\epsilon_{T}^{cb}$ interfere with the SM proportional
to $m_\ell$~\cite{Becirevic:2012jf}. 

\subsubsection{The $B\to \tau\nu$ decay}

The branching fraction of this decay in the SM is given by:
\begin{eqnarray}
\mathcal{B}(B^-\to\tau\bar\nu_\tau)=\tau_{B^-}\,G_F^2 m_\tau^2 f_B^2|V_{ub}|^2\frac{m_B}{8\pi}(1-\frac{m_\tau^2}{m_B^2})^2.
\label{eq:Btotaunu}
\end{eqnarray}
In order to predict the rate in the SM one needs a value for the semileptonic decay
constant of the $B$ meson, $f_B$, and for the CKM matrix element $V_{ub}$. The former is 
calculated in the lattice and the FLAG average of the current results ($N_f=2+1$) is
$f_B=190.5(4.2)$ MeV~\cite{Aoki:2013ldr} while for the latter we use the value resulting 
from the unitarity-triangle fit performed by the CKM-fitter collaboration,
$|V_{ub}|_{\rm CKM}=3.55(16)\times10^{-3}$. With this, we obtain:  
\begin{equation}
\mathcal{B}(B^-\to\tau\bar\nu)^{\rm SM}_{\rm CKM}=0.81(8)\times10^{-4},\label{eq:BtotaunuSM}
\end{equation}
where we have added the errors of $f_B$ and $|V_{ub}|$ in quadratures. The current average 
of the experimental measurements is~\cite{Agashe:2014kda}:
\begin{eqnarray}
\mathcal{B}(B^-\to\tau\bar\nu)^{\rm expt}=1.14(27)\times10^{-4},
\end{eqnarray}
which is compatible (the tension is $1.5\sigma$) with the SM. The measurement however leaves room for NP contributions 
of the type:
\begin{equation}
\epsilon_L^{ub,\tau}-\epsilon_L^{ub,l}\sim 0.2,,\hspace{1cm}l=e,\,\mu, \label{eq:vl13}
\end{equation}
although a LUV combination $\epsilon_{s_R,s_L}^{ub,\tau}$ is also allowed. In any case we want to emphasize
that this tension depends crucially on the value of $|V_{ub}|$ and that one 
needs to bear in mind the long-standing discrepancy between the determinations 
from the inclusive $B\to X_u\ell^+\nu$ and exclusive $\bar B\to M\ell\bar\nu$~\cite{Agashe:2014kda}
decays, $|V_{ub}|_{\rm inc}=4.13(49)\times10^{-3}$~\cite{Amhis:2014hma}
and $|V_{ub}|_{\rm exc}=3.28(29)\times10^{-3}$ respectively. In fact, using the
inclusive value one obtains $\mathcal{B}(B^-\to\tau\bar\nu_\tau)^{\rm SM}_{\rm inc}=1.09(26)\times10^{-4}$. 

\subsubsection{The $t\to\tau\nu q$ decay}
 
An important constraint in the NP scenarios discussed below could come from measurements of 
the semileptonic decay rates of the top quarks into $\tau$. These have been obseved by
CDF, with 2 candidate events where the SM expectation is $1.00\pm0.06\pm0.16$, 
 with $1.29\pm0.14\pm0.21$ events of expected background~\cite{Abulencia:2005et}. This allows to set a bound on
 the ratio:
 \begin{equation}
 R_{t\tau}=\frac{\Gamma(t\to\tau\nu q)}{\Gamma(t\to\tau\nu q)^{\rm SM}}, 
 \end{equation}
 namely, $R_{t\tau}<5.2$ at $95\%$ C.L., or 
\begin{equation}
 (\epsilon_L^{tb,\tau})^*<1.3.\label{eq:vl33exp} 
\end{equation}
Finally, note that this bound is obtained at energy scales of the order of the top-quark mass, so that
the Wilson coefficient needs to be run down to $\mu=m_b$ in order to study the consequences in $B$-meson
decays. Nonetheless, the one-loop anomalous dimensions of the vector and axial currents are zero in QCD and we
neglect the effects of the electroweak contributions.

\section{Model-independent discussion}
\label{SEC3}

As discussed in the previous section, the $R_K$ anomaly can only be accommodated by LUV
contributions to the semileptonic operators $\mathcal{O}^{(\prime)}_{9,10}$. The effect required
is compatible with some of the NP scenarios suggested by the analysis of $B_s\to\mu^+\mu^-$
and $B\to K^*\mu^+\mu^-$, in particular, with NP coupled to left-handed quarks and
left-handed muons,
\begin{eqnarray}
&\delta C_9^\mu=-\delta C_{10}^\mu=-0.5,\nonumber\\
&\delta C_9^e=\delta C_{10}^e=0. \label{eq:lhmuons} 
\end{eqnarray}
Scenarios with right-handed quark currents are disfavored because they worsen 
the agreement with the measured branching ratio of $B_s\to \mu^+\mu^-$, eq.~(\ref{eq:WCBsll}).
Scenarios with right-handed lepton currents do not produce any sizable effect in $R_K$~\cite{Hiller:2014yaa}.

It is important to keep in mind that the tension in $B_s\to\mu^+\mu^-$ is not statistically
very significant and it is not clear yet if the anomalies in $B\to K^*\mu^+\mu^-$ could be caused
by uncontrolled hadronic effects. Thus, the measurement of $R_K$ can be explained, alternatively, 
by a NP scenario coupled predominantly to electrons. The $b\to s e e$ decays are far less constrained
experimentally than their $b\to s \mu \mu$ siblings and all combinations that could be derived 
from~(\ref{eq:WCRK}) are in principle possible. 

Nevertheless, for reasons that will become apparent shortly, in this work we focus
on NP interpretations of $R_K$ where the coupling to electrons is not altered. The required 
left-handed--left-handed contributions to $b\to s \mu \mu$ can only be generated
by the operators $Q^{(1)}_{\ell q}$ and $Q^{(3)}_{\ell q}$ of the SMEFT
Lagrangian. These also contribute to the $b\to s\nu\nu$ transitions and $Q^{(3)}_{\ell q}$  
induces LUV effects in charged-current decays, eq.~(\ref{eq:SMEFTcc}). 
For muon and electrons the experimental data from rare $B$ decays render these effects negligible; 
however rare decays to $\tau$ leptons are poorly 
constrained and the loop-suppression factor characteristic of the neutral-current
transitions in the SM could compensated by a strong flavor hierarchy. This was illustrated
in ref.~\cite{Bhattacharya:2014wla}, where the $R_K$ and $R_{D^{(*)}}$ anomalies
were connected assuming a $Q^{(3)}_{\ell q}$ contribution coupled exclusively to third 
generation of quarks and leptons (in the interaction basis) and generic assumptions on the 
unitary flavor mixing matrices. In fact, this mechanism had been introduced earlier in ref.~\cite{Glashow:2014iga}
to argue that violation of lepton universality would \textit{necessarily} lead to lepton-flavor
violation in $b\to s\ell\ell'$ (semi)leptonic transitions (see also recently~\cite{Boucenna:2015raa}).

\subsection{MLFV}

Given the MFV assumption for the lepton sector and generalizing  eq.~(\ref{CEFTMLFV}) to all orders in the 
Yukawa expansion (see Ref.~\cite{Kagan:2009bn} for a discussion of the quark case), 
the operators singled out above, $Q^{(1)}_{\ell q}$ and 
$Q^{(3)}_{\ell q}$, read:
\begin{equation}
\mathcal{L}^{\rm NP}=\frac{1}{\Lambda^2}\left[(\bar q_L\,C_q ^{(1)}\gamma^\mu q_L)(\bar \ell_L\,F(\hat Y_e\,\hat Y_e^\dagger)\gamma_\mu \ell_L)
+(\bar q_L\,C_q ^{(3)}\gamma^\mu \vec{\tau}q_L)\cdot(\bar \ell_L\,F(\hat Y_e\,\hat Y_e^\dagger)\gamma_\mu\vec{\tau}\ell_L)\right], \label{eq:MILag}
\end{equation}
where $\hat Y_e$ is the charged lepton Yukawa normalized as in eq.~(\ref{hatYe}) and $C_q^{(1,3)}$
are generic $3\times3$ hermitian matrices in quark flavor space. $F(x)$ is a general regular function  
whose zeroth order we neglect, $F(0)=0$, since we are interested in non-trivial flavor effects, and 
it is normalized such that $F^\prime(0)=1$ which can always be done redefining $C_q^{(1,3)}$. For the sake of clarity in the
forthcoming discussion, we assume that the two operators have the same structure in lepton-flavor space. 
Nonetheless, the same conclusions would follow from the more general case.

In the present MLFV set-up, the unitary rotation that takes to the mass basis also diagonalizes
the flavor structure of the NP operators, generating LUV effects governed by the normalized leptonic Yukawa couplings and
without introducing LFV in the process. Thus, the above Lagrangian produces the contributions to $C_9^\alpha$:
\begin{align}
\delta C_9^\alpha=&\left(C_q^{(1)}+C_q^{(3)}\right)_{sb}\,F\left(\frac{m_\alpha^2}{m_\tau^2}\right)\,\frac{4\pi^2v^2 }{e^2\lambda_{ts}\Lambda^2} \\ 
=&\left(\mathcal{O}\left(\frac{m_e^2}{m_\tau^2}\right) \,,\frac{m_\mu^2}{m_\tau^2}+
\mathcal{O}\left(\frac{m_\mu^4}{m_\tau^4}\right) \,,\,f\,\right)\frac{4\pi^2v^2 }{e^2\lambda_{ts}\Lambda^2}\left(C_q^{(1)}+C_q^{(3)}\right)_{sb}\,,\label{eq:MIanamu1}
\end{align}
where $f\equiv F(1)$, $\alpha$ denotes the lepton flavor index, which is expanded as an array in the second line, and the subindex $sb$ denotes the entry in the  $C_q^{(1,3)}$ matrices. 
In this case, the $b\to s\ell\ell$ anomalies would be explained by NP coupled predominantly to muons:
\begin{equation}
\left(C_q^{(1)}+C_q^{(3)}\right)_{sb}\frac{v^2}{\Lambda^2}=\,\left(\frac{m_\tau}{m_\mu}\right)^2\lambda_{ts}\frac{\alpha_{e}}{\pi}\delta C_{9}^\mu
\simeq0.33\,|\lambda_{ts}|\,,\label{eq:MIanamu1}
\end{equation}
where we have applied the scenario in eq.~(\ref{eq:lhmuons}) which, 
for Wilson coefficients of order one, yields an effective NP scale of $\Lambda\simeq2$ TeV.

In order to discuss the consequences of this ansatz in the physics of the tauonic $B$-meson decays,
we first study the simplest case introduced in Sec.~\ref{Sec:MFV} in which  $F(\hat Y_e\hat Y_e^\dagger)=\hat Y_e\hat Y_e^\dagger$ or, equivalently, $f=1$. The most striking consequence of this scenario is the large enhancement
produced in the tauonic transitions as the corresponding operators are multiplied by a large factor. For instance,
for the rare $B_s\to\tau\tau$ and $B\to K\tau^-\tau^+$ decays one is led to: 
\begin{equation}
\overline{\mathcal{B}}_{s\tau}\simeq1\times10^{-3},\hspace{3cm}\mathcal{B}(B\to K\tau^-\tau^+)\simeq2\times10^{-4},
\label{eq:btostautaupredictions}
\end{equation}
These are still an order of magnitude below the bounds obtained from the experimental
limits in eqs.~(\ref{eq:btostautau}), although the predicted boost of $\sim10^3$ in these
decay rates with respect to the SM should be testable in a next round of experiments at Belle II.

A similar enhancement is produced in other operators. In particular, $b\to s\nu^k\bar\nu^l$, where the neutrinos are in the mass basis, receives
a contribution,
\begin{eqnarray}
    \delta C_\nu^{kl}= U^\dagger_{k\tau}\left(C_q^{(1)}-C_q^{(3)}\right)_{sb}U_{\tau l}\frac{4\pi^2 v^2}{e^2\lambda_{ts}\Lambda^2}\,.
\end{eqnarray}
Unlike $b\to s\tau\tau$, this decay is well constrained experimentally; according to  eq.~(\ref{eq:bsnunuexpt}) we have
\begin{equation}
\left(C_q^{(1)}-C_q^{(3)}\right)_{sb}\,\frac{v^2}{\Lambda^2}\lesssim 0.01\,|\lambda_{ts}|, \label{eq:MIanamu4}
\end{equation}
that, in combination with eq.~(\ref{eq:MIanamu1}), gives 
\begin{equation}
\left(C_q^{(1)}-C_q^{(3)}\right)_{sb}\lesssim 0.03\,\left(C_q^{(1)}+C_q^{(3)}\right)_{sb}
\,, \label{eq:MIanamu5}
\end{equation}
which effectively sets the constraint $C_q^{(1)}=C_q^{(3)}$. 
Although eq.~(\ref{eq:MIanamu5}) seems to impose a fine-tuning, we will see in Sec.~\ref{Sec:lepqtoquarkillustrated} 
how the relation $C_{q\ell}^{(1)}=C_{q\ell}^{(3)}$ can arise in a specific model from the quantum numbers
for the new particles.

\begin{figure}[t]
\begin{tabular}{cc}
  \includegraphics[width=65mm]{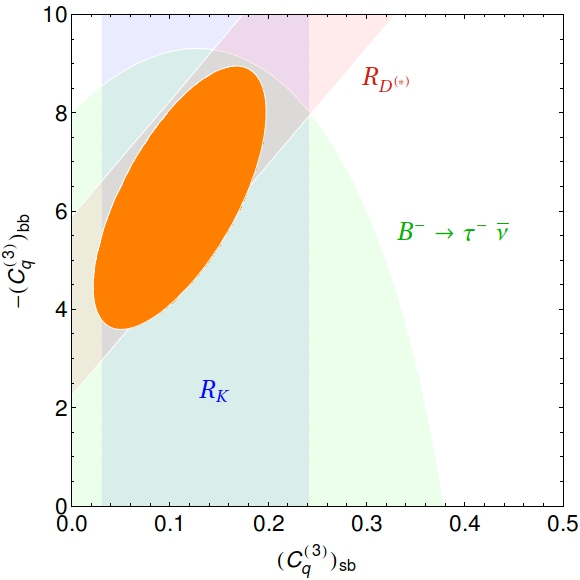}\hspace{0.8cm} & \hspace{0.8cm}  \includegraphics[width=62mm]{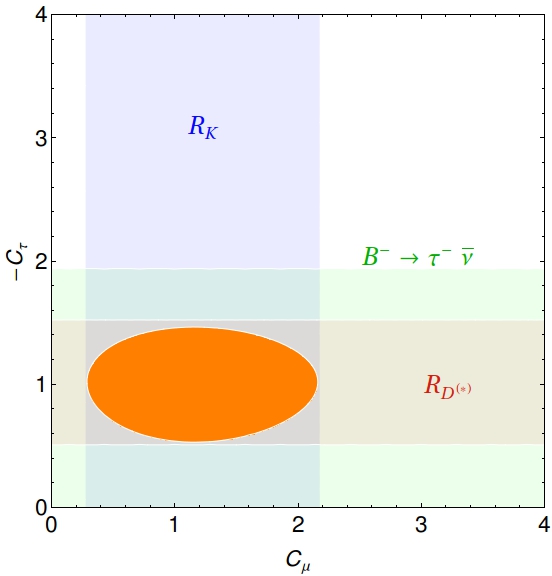} 
\end{tabular}
\caption{Constraints at 90\% CL on the Wilson coefficients of the SMEFT as defined in eq.~(\ref{eq:MILag}) in the two MLFV scenarios
discussed in the text. \textit{Left-hand:} Scenario in which flavor breaking in the lepton
sector is given by $\hat Y_e\,\hat Y_e^\dagger$ and where we have used $\Lambda=1$ TeV as a reference
for the NP scale. \textit{Right-hand:} Scenario with lepton-flavor breaking given by the function
$F(\hat Y_e\,\hat Y_e^\dagger)$, where $\Lambda=3$ TeV and using as labels $C_\mu=(C^{(3)}_q)_{sb}$ and $C_\tau=f\,(C^{(3)}_q)_{sb}$.
\label{Fig:CPMLFV}}
\end{figure}

There is another modification in the charged-current effective Lagrangian, eq.~(\ref{eq:SMEFTcc}).
Neglecting for simplicity the $k=1$ flavor entry one finds that all these decays are modified by
the combination: 
\begin{equation}
\epsilon_L^{ib,\tau}=-\frac{v^2}{\Lambda^2}\left(\frac{V_{is}}{V_{ib}}\left(C_q^{(3)}\right)_{sb}+\left(C_q^{(3)}\right)_{bb}\right).\label{eq:MIanamu7}
\end{equation}
The first term is the same entering in $R_K$, eq.~(\ref{eq:MIanamu1}), once the constraint from $b\to s\nu\bar\nu$, eq.~(\ref{eq:MIanamu5}),
is taken into account. The second term is double-CKM suppressed and if $(C_q^{(3)})_{bb}$ is of the same order of
magnitude as $(C_q^{(3)})_{sb}$, then its contribution is negligible and the correction to the charged current (semi)leptonic $B$ tauonic
decays is entirely given by the one required to understand the $b\to s\ell\ell$ anomalies. For example,
in $B\to D^{(*)}\tau\nu$ one obtains that $\epsilon_L^{cb,\tau}=-0.16$. This has the right
size but the opposite sign necessary to explain $R_{D^{(*)}}$, eq.~(\ref{eq:vl23expt}), producing a
deficit of tauonic decays with respect to the electronic and muonic ones instead of the excess observed experimentally.
The same effect appears in the $b\to u \tau^-\bar\nu$ transition, $\epsilon_L^{ub,\tau}=-0.16$, leading to a similar conflict with
the experimental rate of $B^-\to\tau^-\bar\nu$, eq.~(\ref{eq:vl13}).

A first strategy to solve this problem is to introduce a hierarchy in the quark flavor
 structure such that $-V_{cb}\,(C_q^{(3)})_{bb}\gg (C_q^{(3)})_{sb}$.
Another
solution is to re-introduce the generic function $F(Y_eY_e^\dagger)$ such that  $f\simeq-1$. In this
case one can neglect the contribution from $(C_q^{(3)})_{bb}$ and explain simultaneously the $b\to s\ell\ell$
and tauonic $B$-decay anomalies without demanding any hierarchy among the effective parameters. Note that 
in this scenario the constraint obtained from the decays into neutrinos, eq.~(\ref{eq:MIanamu5}), and the
prediction of the strong enhancement of the tauonic decay rates in eq.~(\ref{eq:btostautaupredictions}) hold.
\footnote{One could also investigate the effect of $(C_q^{(3)})_{db}$ in the
charged-current $B$-decays. However, this does not improve significantly the description 
because the $b\to d\ell\ell$ data is still scarce and the fact that this
term leads to contributions with different CKM suppression in the $b\to c\ell\bar\nu$ and $b\to u\ell\bar\nu$ decays.}

In figs.~\ref{Fig:CPMLFV} we show the contour plots given by the different experimental results
in the parameter space of these two scenarios. On the left panel we have the case in which $f=1$
and where we have chosen $\Lambda=1$ TeV as the effective NP mass. As we can see, all the measurements 
discussed above can be accommodated, although at the price of making $(C_q^{(3)})_{bb}$ large and close 
to the nonperturbative limit $(C_q^{(3)})_{bb}\sim\mathcal{O}(4\pi)$. An important limit to 
$(C_q^{(3)})_{bb}$ could come from the $t\to q\tau\nu$ decay. The current bound in eq.~(\ref{eq:vl33exp})
translates into $-54<-(C_q^{(3)})_{bb}<21$ (using $\Lambda=1$ TeV) which is still a factor two above 
the relevant region. A modest improvement of this bound could probe this scenario thoroughly. For example, 
an improvement of a factor 4 over the CDF measurement, $R_{t\tau}<1.3$, would result in $-(C_q^{(3)})_{bb}<2.24$.

On the right panel, we show the scenario where $f\simeq-1$, for an effective scale of $\Lambda=3$ TeV
and using as labels $C_\mu=(C^{(3)}_q)_{sb}$ and $C_\tau=f\,(C^{(3)}_q)_{sb}$. This is an interesting hypothesis
to explain naturally the various anomalies with short distant physics in the few-TeV range, especially because
these involve sizable effects in processes which span different degrees of suppression in the SM.
In particular, the neutral-current transition in $R_K$ is loop-suppressed with respect to the charged-current,
tree-level ones, $B\to D^{(*)}\tau\nu$ and $B^-\to\tau^-\bar\nu$. In our scenario, the difference between the 
apparent NP effective scales in these processes is explained by the hierarchy in the couplings introduced
by the different lepton masses, with $\alpha_{e}/\pi\sim (m_\mu/m_\tau)^2$. 

\begin{figure}[t]
\begin{center}
\includegraphics[scale=0.32]{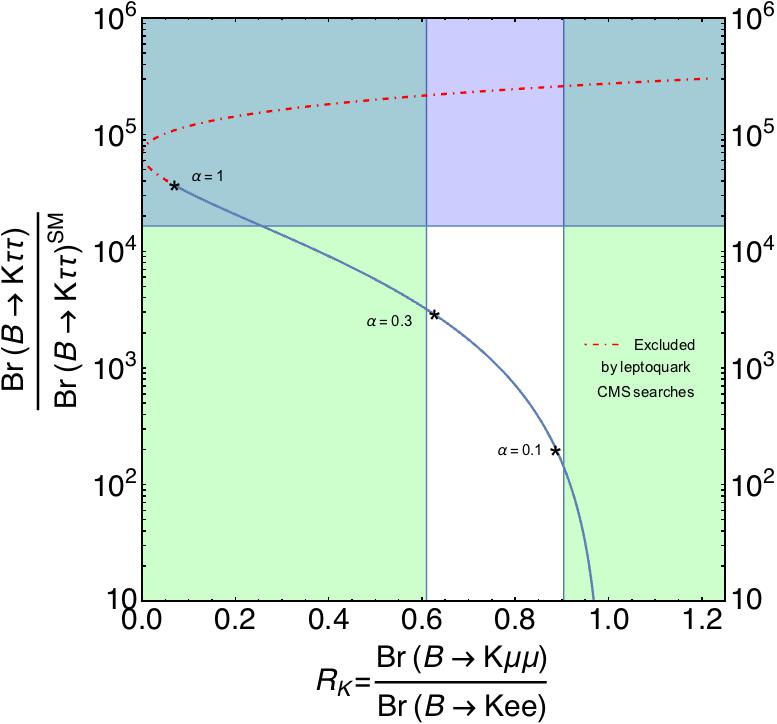}
\caption{
Prediction of MLFV to first order in lepton Yukawas in the form of a line in the plane of $R_K$ \textit{vs} $\mathcal{B}(B\to K\tau\tau)$, the experimentally
allowed region at $90\%$ CL is the white band. In the case in which the effect is produced by a hypercharge-2/3 leptoquark and
applying MFV in the quark sector too, CMS direct searches exclude the red dashed part of the line. The points marked with $\alpha\equiv g^2/(4\pi)=1,0.3,0.1$ correspond to the prediction
for a $600$ GeV leptoquark with the coupling constant defined in eq.~(\ref{TMFV}). 
\label{Fig:CPMFV}}
\end{center}
\end{figure}

Finally, let us discuss the case in which MFV is imposed also in the quark sector. Let us assume for simplicity
the scenario in eq.~(\ref{CEFTMFV}). In this case, the $b\to s\ell\ell$ anomalies are explained with  
\begin{equation}
\left(c_q^{(1)}+c_q^{(3)}\right)\frac{v^2}{\Lambda^2}=\,\left(\frac{m_\tau}{m_\mu}\right)^2\frac{\alpha_e}{\pi}\delta C_{9}^\mu,\label{eq:MIMFVanamu1}
\end{equation}
where the flavor structure in the quark sector is given by $\,\lambda_{ts}$. Note that in this case, there 
is no CKM suppression of the SM contribution with respect to the nonstandard one, so that:
\begin{equation}
\left(c_q^{(1)}+c_q^{(3)}\right)\frac{v^2}{\Lambda^2}=0.33.\label{eq:MIMFVanamu2}
\end{equation}
Therefore, the effective mass should be close to the electroweak scale, and perturbative couplings $c_q^{(1)}+c_q^{(3)}$
 are only possible for a new physics scale below $1.5$ TeV.
In this approach one obtains the same predictions for the tauonic channels presented in eq.~(\ref{eq:btostautaupredictions})
and the constraint $c_q^{(1)}=c_q^{(3)}$ after considering the decays into neutrinos. However, 
using eqs.~(\ref{eq:SMEFTcc},\ref{CEFTMFV}) we find that the contributions to the charge-current $B$ decays now are:
\begin{equation}
\epsilon^{ib}_L\simeq-\frac{v^2}{\Lambda^2}\,c_q^{(3)}\frac{y_i^2}{y_t^2},\label{eq:MIMFVcc}
\end{equation}
such that they are suppressed by small up-quark Yukawas and negligible. 

Finally, in fig.~\ref{Fig:CPMFV} we graphically display the correlation between $R_K$ and $B\to K\tau\tau$ for $f=1$, that is, the case 
in which we keep the leading term only in the expansion in the leptonic Yukawas. In this case there is only one 
NP parameter controlling both processes, a variation of which produces the curve shown. Allowed experimental
values at $90\%$ CL correspond to the white region, and one can see that accommodating $R_K$ leads to a $\mathcal{B}(B\to K \tau\tau)$ that is a few $\times 10^3$ 
larger than the SM value. If the effect is produced by a leptoquark, then CMS bounds rule out part of the line, see the Sec.\ref{Sec:lepqtoquarkillustrated} for details.

\section{A Leptoquark Model}
\label{Sec:lepqtoquarkillustrated}

The leptoquark particles that couple to SM fermions via operators of dimension $d\leq4$  are either
spin-0 or spin-1 bosons and they can be sorted out in terms of their quantum numbers 
(see~\cite{Buchmuller:1986zs,Davidson:1993qk,Hewett:1997ce,Sakaki:2013bfa} and the appendix for details).
There is a total of 5 scalars and 5 vector bosons as shown in tabs.~\ref{TabS} and~\ref{TabV}.  Assuming that their masses
are above the electro-weak scale, we compute their contribution to the 4-fermion operators of the SMEFT in tab.~\ref{tab:matching}.
In this table one can see that only 5 of the scalar and 3 of the vector boson leptoquarks contribute to $B$-physics and
the coefficients they produce for the operators in the low-energy Lagrangian of eqs.~(\ref{eq:semilOps},\ref{eq:nonSM:scalars},\ref{eq:Leffcc})
are given in Tab~\ref{Tab:LQbsll}.

The number of independent operators that enter neutral-current $B$ (semi-)leptonic decays is, after imposing 
the full $SU(2)_L\times U(1)_Y$ symmetry, 6, whereas there are 4 charged current operators in 
eq.~(\ref{eq:Leffcc}) which receive contributions from leptoquark models. There is therefore a priori 
enough potential experimental inputs to non-trivially test the hypothesis of a leptoquark in $B$-physics.

The crucial test for these models however would be the detection of the leptoquark resonances. Since they carry color, the LHC is a 
powerful tool in the search for leptoquarks, which has however yielded only bounds so far, pushing the mass scale to the TeV range~\cite{Khachatryan:2014ura,Khachatryan:2015bsa}.

Using Table~\ref{Tab:LQbsll} and  the  previous EFT study of
the experimental data, it is straight-forward to select
the leptoquark model that would better fit the data: a hyper-charge $2/3$, $SU(2)_L$-singlet, color-fundamental
vector boson. The Lagrangian reads:
\begin{eqnarray}
\mathcal{L}_{V}&=&(V_{-2/3}^\mu)^\dagger\left(D^2+M^2 \right)(V_{-2/3})_\mu +\left(g_{\ell q}\,\bar \ell_L\gamma_\mu q_L+g_{ed}\,\bar e_R\gamma_\mu d_R\right)\,V^\mu_{-2/3} +{\rm h.c.}, \label{eq:LagV}
\end{eqnarray}
where gauge and flavor indices have been omitted. Note that this model avoids the contributions
to $C_{\nu}$ since the $SU(2)_L$ contraction  only couples up quarks to neutrinos and
down quarks to electrons $\bar q_L \ell_L=\bar u_L\nu_L+\bar d_L e_L$, hence $C_q^{(1)}=C_q^{(3)}$.
 Also note that this model generates the chiral structure for the semi-leptonic operators suggested by data.

The flavor structure is
the decisive part of the model and the focus of this work. We will use the MFV hypothesis, which was studied in the context
of leptoquarks in Ref.~\cite{Davidson:2010uu}. Here we will implement our hypothesis in two ways:

\begin{itemize}
\item{\bf Minimal Lepton Flavor Violation}

If we formally impose only the flavor symmetry of the leptons, $SU(3)_\ell \times SU(3)_e\times U(1)_L\times U(1)_{e-\ell}$, we have
that $g_{q\ell} V_{-2/3}$ should transform as a $(3,1)_{1,-1}$ and $g_{ed} V_{-2/3}$ as $(1,3)_{1,1}$. MLFV prescribes that
$g_{q\ell,ed}$ should be built out of Yukawas, in this case $Y_e=\varepsilon_e\hat Y_e$. Triality, which is the conservation of fundamental
indices modulo 3~\cite{Georgi:1982jb}, prevents building a $(3,1)$ or $(1,3)$ representation from any number of Yukawas $(3,\bar 3)$.
\footnote{A combination of $n$ $Y_e$'s (with triality $(1,2)$) and $m$ $Y_e^\dagger$'s (with triality $(2,1)$) 
has triality ($n+2m$ mod $3$, $m+2n$ mod $3$). Although it looks like the two final trialities are unrelated, the sum of the two is always
$0$ mod $3$; this implies that only the trialities $(0,0), (1,2), (2,1)$ are possible and remember that a representation $(3,1)$ has triality $(1,0)$.}
This means that we have to assign flavor to $V_{-2/3}$, the simplest choice
 being a fundamental of either $SU(3)$ flavor group. Of the possible choices, the one that yields unsuppressed LUV in
the $Q_{\ell q}$ operators is:
\begin{align}
V^\mu_{-2/3}&\sim (3,1)_{1,-1}\,, & 
g_{\ell q}&=g_q \hat Y_e\,, 
& g_{de}&=g_d\varepsilon_e^*\,,
\end{align}
where $g_{q,d}$ have a quark flavor index but no lepton index and we neglect higher powers in Yukawas. 
There is an interesting alternative to this scenario that however leads to the same low energy Lagrangian.
Indeed, one might object that the above model inserts Yukawa couplings as prescribed by MFV 
but does not justify how those Yukawas got there in the first place, and is in this sense incomplete.
A solution to this is the gauge flavor symmetry scenario~\cite{Grinstein:2010ve}. In this case the Yukawas
are the vev of the inverse of some scalar fields, 
$\mathcal Y_e$ that do transform ($\mathcal{Y}_e\sim(\bar 3\,,3)$) under the gauged 
flavor group: $Y_e\propto 1/\left\langle \mathcal Y_e\right\rangle$. The  Lagrangian would be, choosing $V_{-2/3}\sim (3,1)$:
\begin{eqnarray}
\mathcal{L}_{V}&=&(V_{-2/3}^\mu)^\dagger \left(D^2+M^2+\lambda_e\, \mathcal Y_e^\dagger \mathcal Y_e \right)(V_{-2/3})_\mu +g_{ q}\bar \ell_L\gamma_\mu q_L\,V^\mu_{-2/3} +{\rm h.c.}, \label{eq:LagV}
\end{eqnarray}
Note that the coupling to $\bar e_R\gamma_\mu d_R$ requires an irrelevant operator.
If $M^2$ is negligible with respect to $\mathcal Y_e$, a case in which all operators would be marginal and
the theory classically conformal, the effective operator $Q_{\ell q}^{(1)}$ has a coefficient
$g_{q}g_{q}^*/\left\langle \mathcal Y_e^\dagger \mathcal Y_e \right\rangle\propto Y_eY_e^\dagger$, 
just as in eq.~(\ref{CEFTMLFV}). Such a model however faces the more pressing question of renormalizability, 
due to the presence of massive vector bosons. In this sense one can postulate a strong sector in the spirit of QCD that yields
as a composite state the vector boson $V_{-2/3}^\mu$ as a ``$\rho$'' particle, whereas the flavor structure is dictated by the
short distance physics of a gauged flavor symmetry  model. Also note that a scalar leptoquark would yield a renormalizable theory.

In both cases, at low energy the coefficients of the operators are, neglecting the coupling $g_d\varepsilon_e$:
\begin{equation}
 \frac{\alpha_e}{\pi}\lambda_{ts}\delta C_9=-\frac{v^2}{M^2}\left(\frac{m_\mu}{m_\tau}\right)^2({g}_q^ {s})^*{g}_q^b
\end{equation}
where $g_i$ is $i$-th component of the quark-family vector $g_{\ell q}$, we have already made the rotation to the down-type
quark mass basis and $\delta C_{10}=-\delta C_{9}$. The modification to the charged current Lagrangian is:
\begin{eqnarray}
\epsilon_L^{kj,\tau}=\frac{1}{2}\frac{v^2}{M^2}\sum_k\frac{V_{ik}}{V_{ij}}({g}_q^k)^*{g}^j_q\,.
\end{eqnarray}

In fig.~\ref{fig:PhenoLQ} we show the experimental constraints on the plane of the 
(real) leptoquark couplings $g_q^s$ and $g_q^b$ and using $M=0.75$ TeV. The gray band corresponds to the perturbativity bound $g_q^i=\sqrt{4\pi}$.
Finally, this model has to confront direct searches at the LHC. Searches for vector leptoquarks decaying to a $b$ quark and a
$\tau$ lepton have been done by CMS, setting a limit on the mass  $M>600$GeV~\cite{Khachatryan:2014ura,Khachatryan:2015bsa}.\footnote{In 
deriving this bound we have used  branching ratios to $b \tau$ and $t \nu_\tau$ of $50\%$.} 
\begin{figure}[t]
\includegraphics[scale=0.4]{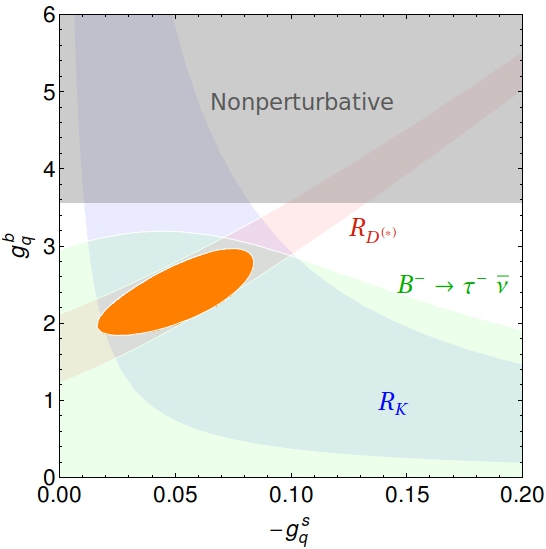}

\caption{\label{fig:PhenoLQ}
Constraints at 90\% CL on the plane of the (real) leptoquark couplings $g_q^s$ and $g_q^b$ and using $M=0.75$ TeV. The gray
band corresponds to the perturbativity bound $g_q^i=\sqrt{4\pi}$.}
\end{figure}

\item{\bf Quark and Lepton Minimal Flavor Violation.}

If we now consider the whole flavor group for both quarks and leptons, the number of free parameters decreases
since the quark flavor structure is now dictated by the up type Yukawas.
Like in the previous case, triality implies that $V^{\mu}_{2/3}$ has to transform under the flavor group.
We will write the quark flavor group as $U(3)^3=SU(3)_q\times SU(3)_u\times SU(3)_d\times U(1)_B\times U(1)_{u-q}\times U(1)_{d-q}$. 
The choice of flavor transformations that yields the operator $Q_{\ell q}$ 
with QFV and LUV is:
\begin{align}\label{TMFV}
V_{-2/3}&\sim (1,3)_{1,-1}\times(1,3,1)_{1/3,1,1}\,, & 
g_{q\ell}&=g\hat Y_u^\dagger \otimes \hat Y_e\,,
& g_{de}&=g^\prime \varepsilon_e^*\varepsilon_d \hat Y_u\otimes \hat Y_d^\dagger\,,
\end{align}
where now $g,g^\prime$ are overall flavorless constants and note that the $U(1)$ charge assignment is consistent with hypercharge. 
This dictates that the coefficients of the operators are, neglecting $g^\prime \varepsilon_e^*\varepsilon_d$ :
\begin{equation}
 \frac{\alpha_e}{\pi}\delta C_9=-\frac{v^2}{M^2}\left(\frac{m_\mu}{m_\tau}\right)^2\left|g\right|^2 .
 \end{equation}
and $\delta C_{10}=-\delta C_{9}$. Note that the sign of the contribution is fixed and  $g\sim 1$ implies  $M\sim 600$ GeV, which is
around the current experimental CMS limit.
The MLFV prediction in fig.~\ref{Fig:CPMFV} depends on the combination $g/M$ for this model. The points marked as $\alpha=g^2/(4\pi)=0.1,0.3,1$ correspond to a mass $M=600$ GeV.
The red dashed part of the line has $\alpha\geq1$ and we consider it excluded.

\end{itemize}

\section{Conclusions}

We have analyzed various anomalies in the neutral- and charged-current (semi)leptonic $B$-meson decays 
that suggest the presence of new interactions violating lepton universality. Although this leads to the expectation of sizable lepton flavor violation in $B$ decays 
(as discussed recently and abundantly in the literature), we have shown that this is not the case in
a general class of new-physics scenarios with minimal flavor violation. Namely, in case one can neglect
flavor effects from the neutrino mass generation mechanism, one finds that charged-lepton flavor is 
preserved but universality is not, with the violation of the latter being controlled by charged-lepton masses.
In these scenarios, the skewed ratio $R_K$ found experimentally is explained by new physics coupled
more strongly to muons than to electrons. Furthermore, the tauonic $B$-decays receive a strong 
enhancement due to the relative factor $(m_\tau/m_\mu)^2$ in the corresponding couplings to the leptons.

We have first explored the phenomenological consequences of this hypothesis at the level of the effective
operators of the standard model effective field theory and have selected a linear combination of them involving only $SU(2)_L$ doublets
as the most plausible explanation for all the anomalies. Accommodating $R_K$ in this scenario implies an $\mathcal{O}(10^3)$ boost,
with respect to the standard model, of the $B_s\to\tau\tau$, $B\to K^{(*)}\tau\tau$ and $B\to K\nu\bar\nu$ decay
rates. The predicted rates to charged $\tau$ leptons are an order of magnitude below the current experimental
limits and they could be tested in future experiments at $B$ factories. The decay into neutrinos, which is much
better measured, poses a strong constraint on the new physics that can be accounted for naturally if 
it does not couple neutrinos and down-type quarks. As for charged current decays, 
the enhancement for tauonic decays is approximately the same enhancement that the standard model presents
due to charged current decays occurring at tree level as opposed to 1-loop rare decays. In fact, a remarkable outcome
is that $R_K$, $R_{D^{(*)}}$ and $B^-\to\tau^-\bar\nu$ anomalies can be explained simultaneously and naturally with
new-physics effective mass in the multi-TeV range. A manifest advantage of using an effective field theory setup is that it shows a 
model-independent interplay between the $B$-decays of interest and top-physics, with the $t\to q\nu \tau $ turning
to be a complementary and powerful way to test these scenarios.

Finally, as an illustration of our hypothesis in model building, we have surveyed the contributions of all possible
spin-0 and spin-1 leptoquark particle models to $B$-decays. We have first projected their contributions into 
the effective operators of the standard model effective field theory, integrating the leptoquark fields out,
and then we have applied the conclusions of our study to select a unique model that is better suited to fit
the data (even if there is room for other possibilities):
an $SU(2)_L$ singlet, color-fundamental vector boson of hypercharge $2/3$. It was shown how the assumed flavor structure of
minimal lepton flavor violation could arise from a Lagrangian with operators of dimension~$\leq4$, and a particular case was presented
in which the flavor structure arose from the vev of scalar fields.

\section{Acknowledgments}

This work was supported in part by DOE grants DE-SC000991 and DE-SC0009919. JMC has received funding 
from the People Programme (Marie Curie Actions) of the European Union's 
Seventh Framework Programme (FP7/2007-2013) under REA grant agreement
n PIOF-GA-2012-330458 and acknowledges the Spanish Ministerio de Econom\'ia y 
Competitividad and european FEDER funds under the contract FIS2011-28853-C02-01
for support. 

\appendix
 
\section{Leptoquark Bosons}
\label{AppLQ}

\begin{sidewaystable}[h]
\begin{center}
\begin{tabular}{|c|cccccccccc|}
\hline
LQ&$C_{\ell q}^{(1)}$&$C_{\ell q}^{(3)}$&$C_{\ell d}$&$C_{qe}$&$C_{ed}$&$C_{\ell edq}$&$C_{lequ}^{(1)}$&$C_{lequ}^{(3)}$&$C_{eu}$&$C_{\ell u}$\\
\hline
$\Delta_{1/3}$&$\frac{y_{\ell q}^{\beta i,A}(y_{\ell q}^{\alpha j,A})^*}{4M^2}$&$-\frac{y_{\ell q}^{\beta i,A}(y_{\ell q}^{\alpha j,A})^*}{4M^2}$&0&0&0&0
&$-\frac{y_{eu}^{\beta i, A}(y_{\ell q}^{\alpha j, A})^*}{2M^2}$&$\frac{y_{eu}^{\beta i, A}(y_{\ell q}^{\alpha j, A})^*}{8M^2}$&$\frac{y_{eu}^{\beta i, A}(y_{eu}^{\alpha j, A})^*}{2M^2}$&0\\
$\vec{\Delta}_{1/3}$&$\frac{3y_{\ell q}^{\prime\beta i,A}(y_{\ell q}^{\prime\alpha j,A})^*}{4M^2}$&$\frac{y_{\ell q}^{\prime\beta i,A}(y_{\ell q}^{\prime\alpha j,A})^*}{4M^2}$&
0&0&0&0&0&0&0&0\\
$\Delta_{7/6}$&0&0&0&$-\frac{y_{eq}^{\alpha i,A} (y_{eq}^{\beta j,A})^*}{2M^2}$&0&0&$-\frac{y_{\ell u}^{\alpha i, A} (y_{eq}^{\beta j, A})^*}{2M^2}$&$-\frac{y_{\ell u}^{\alpha i, A}(y_{eq}^{\beta j, A})^*}{8M^2}$&
0&$-\frac{y_{\ell u}^{\alpha i,A} (y_{\ell u}^{\beta j,A})^*}{2M^2}$\\
$\Delta_{1/6}$&0&0&$-\frac{y_{\ell d}^{\alpha i\,A}(y_{\ell d}^{\beta j,A})^*}{2M^2}$&0&0&0&0&0&0&0\\
$\Delta_{4/3}$&0&0&0&0&$\frac{y_{ed}^{\beta i\,A}(y_{ed}^{\alpha j,A})^*}{2M^2}$&0&0&0&0&0\\
\hline
$V^\mu_{2/3}$&$-\frac{g_{\ell q}^{\alpha i,A}(g_{\ell q}^{\beta j,A})^*}{2M^2}$&$-\frac{g_{\ell q}^{\alpha i,A}(g_{\ell q}^{\beta j,A})^*}{2M^2}$&
0&0&$-\frac{g_{ed}^{\alpha i, A}(g_{ed}^{\beta j, A})^*}{M^2}$&$\frac{2g_{\ell q}^{\alpha i, A}(g_{ed}^{\beta j, A})^*}{M^2}$&0&0&0&0\\
$\vec{V}^\mu_{2/3}$&$-\frac{3g_{\ell q}^{\prime\alpha i,A}(g_{\ell q}^{\prime\beta j,A})^*}{2M^2}$&$\frac{g_{\ell q}^{\prime\alpha i,A}(g_{\ell q}^{\prime\beta j,A})^*}{2M^2}$&
0&0&0&0&0&0&0&0\\
$V^\mu_{5/6}$&0&0&$\frac{g_{\ell d}^{\alpha j,A} (g_{\ell d}^{\beta i,A})^*}{M^2}$&$\frac{g_{eq}^{\alpha j,A} (g_{eq}^{\beta i,A})^*}{M^2}$&
$\frac{2g_{\ell d}^{\alpha j, A}(g_{eq}^{\beta i, A})^*}{M^2}$&0&0&0&0&0\\
$V^\mu_{5/3}$&0&0&0&0&0&0&0&0&$-\frac{g_{eu}^{\alpha i\,A}(g_{eu}^{\beta j,A})^*}{M^2}$&0\\
$V^\mu_{1/6}$&0&0&0&0&0&0&0&0&0&$\frac{g_{\ell u}^{\alpha j\,A}(g_{\ell u}^{\beta i})^*}{M^2}$\\
\hline
\end{tabular}
\end{center}
\caption{Tree-level leptoquark contributions to the sixth-dimensional four-fermion operators of the SMEFT with
flavor indexes $(Q_{\ell q}^{(1)})^{\alpha\beta,ij}=(\overline q^j \gamma^\mu q_L^i)(\bar \ell^\alpha \gamma_\mu \ell^\beta)$ and
so forth. The internal leptoquark index, $A$, is summed over.
\label{tab:matching}}\vspace{3mm}
\begin{center}
\begin{tabular}{|c | c c c c  c c | c c |}
\hline
LQ & $C_9$ & $C_{10}$ & $C_9^\prime$ &$C_{10}^\prime$ & $C_S$ & $C_S^\prime$ &$C_\nu$ & $C_\nu^\prime$\\
\hline
$\Delta_{1/3}$ &0&0&0&0&0&0&$\frac{1}{2}y_{\ell q}^{\beta b,A}(y_{\ell q}^{\alpha j,A})^*$&0\\
$\vec{\Delta}^\prime_{1/3}$ & $y_{\ell q}^{\prime\beta b,A}(y_{\ell q}^{\prime\alpha j,A})^*$&
$-y_{\ell q}^{\prime\beta b,A}(y_{\ell q}^{\prime\alpha j,A})^*$& 0& 0& 0 &0&$\frac{1}{2}y_{\ell q}^{\prime\beta b,A}(y_{\ell q}^{\prime\alpha j,A})^*$&0\\
$\Delta_{7/6}$ & $-\frac{1}{2} y_{eq}^{\alpha b,A}(y_{eq}^{\beta j,A})^*$ & $-\frac{1}{2} y_{eq}^{\alpha b,A}(y_{eq}^{\beta j,A})^*$ & 0& 0 & 0&0&0&0\\
$\Delta_{1/6}$ & 0& 0& $-\frac{1}{2} y_{\ell d}^{\alpha b,A}(y_{\ell d}^{\beta j,A})^*$ & $\frac{1}{2} y_{\ell d}^{\alpha b,A}(y_{\ell d}^{\beta j,A})^*$ & 0&0&0&$-\frac{1}{2} y_{\ell d}^{\alpha b,A}(y_{\ell d}^{\beta j,A})^*$\\
$\Delta_{4/3}$ & 0& 0& $\frac{1}{2} y_{e d}^{\beta b,A}(y_{e d}^{\alpha j,A})^*$ & $\frac{1}{2} y_{e d}^{\beta b,A}(y_{e d}^{\alpha j,A})^*$ & 0&0&0&0\\
\hline
$V^\mu_{2/3}$ & $-g_{\ell q}^{\alpha b,A}(g_{\ell q}^{\beta j,A})^*$ &$g_{\ell q}^{\alpha b,A}(g_{\ell q}^{\beta j,A})^*$ &
$-g_{e d}^{\alpha b,A}(g_{e d}^{\beta j,A})^*$ &$-g_{e d}^{\alpha b,A}(g_{e d}^{\beta j,A})^*$ & $2 g_{\ell q}^{\alpha b,A}(g_{ed}^{\beta j,A})^*$&$2 (g_{\ell q}^{\beta j,A})^*g_{ed}^{\alpha b,A}$&0&0\\
$\vec{V}^\mu_{2/3}$ & $-g_{\ell q}^{\prime\alpha b,A}(g_{\ell q}^{\prime\beta j,A})^*$ &$g_{\ell q}^{\prime\alpha b,A}(g_{\prime\ell q}^{\beta j,A})^*$ &0&0&0&0&$-2\,g_{\ell q}^{\prime\alpha b,A}(g_{\ell q}^{\prime\beta j,A})^*$&0\\
$V^\mu_{5/6}$ & $g_{eq}^{\alpha j,A}(g_{e q}^{\beta b,A})^*$ & $g_{eq}^{\alpha j,A}(g_{e q}^{\beta b,A})^*$ & $g_{\ell d}^{\alpha j,A}(g_{\ell d}^{\beta b,A})^*$ &
$-g_{\ell d}^{\alpha j,A}(g_{\ell d}^{\beta b,A})^*$&$2g_{\ell d}^{\alpha j,A}(g_{e q}^{\beta b,A})^*$&$2(g_{\ell d}^{\beta b,A})^*g_{e q}^{\alpha j,A}$&0&$g_{\ell d}^{\alpha j,A}(g_{\ell d}^{\beta b,A})^*$\\
\hline
\end{tabular}
\end{center}
\caption{Contributions of the different leptoquarks to the $b\to d_j\ell_\alpha\ell_\beta$ operators with $j=d,s$ and $\alpha, \beta=e,\mu,\tau.$ The normalization $4\pi^2/(e^2\lambda_{ts})\,v^2/M^2$ has
been factored out and $A$ is an internal leptoquark index summed over.\label{Tab:LQbsll}}
\end{sidewaystable}

For the new leptoquark boson to couple to SM particles through dimension $\leq 4$ operators its spin must be  0 or 1,
and the interactions with SM fields described by:
\begin{eqnarray}
\mathcal{L}_{\Delta}&=&\left(y_{\ell u}\,\bar \ell_L\,u_R+y_{eq}\,\bar e_R \,i\tau_2\,q_L\right)\,\Delta_{-7/6}
+y_{\ell d}\,\bar\ell_L\,d_R\,\Delta_{-1/6}
+\left(y_{\ell q}\,\bar\ell^c_L i\tau_2\,q_L+y_{eu}\bar e_R^c\,u_R\right)\,\Delta_{1/3}\nonumber\\
&+&y_{ed}\bar e^c_R\,d_R\,\Delta_{4/3}+y_{\ell q}^\prime\,\bar \ell^c_L i\tau_2\vec{\tau}q_L\,\cdot\vec{\Delta}^\prime_{1/3}+{\rm h.c.},\label{eq:LagS}\\
\mathcal{L}_{V}&=&\left(g_{\ell q}\,\bar \ell_L\gamma_\mu q_L+g_{ed}\,\bar e_R\gamma_\mu d_R\right)\,V^\mu_{-2/3}
+g_{eu}\,\bar e_R\gamma_\mu u_R\,V^\mu_{5/3}+g_{\ell q}^\prime\,\bar \ell_L\gamma_\mu\vec{\tau} q_L\,\cdot\vec{V}^{\prime\mu}_{-2/3}\nonumber\\
&+&\left(g_{\ell d}\,\bar\ell_L\gamma_\mu d_R^c+g_{eq}\,\bar e_R\gamma_\mu q^c_L\right)\,V_{-5/6}^\mu
+g_{\ell u}\,\bar\ell_L\gamma_\mu u_R^c\,V^\mu_{1/6}+{\rm h.c.}, \label{eq:LagV1}
\end{eqnarray}
where $SU(2)_L$ and flavor indices have been omitted, each leptoquark 
is labeled by its hypercharge and $\Delta$ and $V$ denote scalars and vector boson respectively. The SM charge assignments corresponding to each case are
displayed in tabs.~(\ref{TabS}) and~(\ref{TabV}).

\begin{figure}[h]
\begin{minipage}{.5\textwidth}
\begin{tabular}{c | c c c}
Bilinear $(J)$ & $SU(3)_c$ & $SU(2)_L$ &$U(1)_Y$\\
\hline
$\overline \ell_L^cq_L$& $\bar 3$ & 1\,,3 &1/3\\
$\overline e_R^cu_R$& $\bar 3$ & 1 &1/3\\
$\bar\ell_Lu_R$& $\bar 3$ & 2 &-7/6\\
$\overline e_Rq_L$& $\bar 3$ & 2 &-7/6\\
$\bar\ell_Ld_R$& $\bar 3$ & 2 &-1/6\\
$\overline e_R^cd_R$& $\bar 3$ & 1 &4/3\\
\end{tabular}
\caption{\label{TabS}Charge assignment for leptoquark scalars, $\Delta$,\,\,\,\,\,\,\,
as a function of the SM fermion current to which they couple.}
\end{minipage}\hfill
\begin{minipage}{.5\textwidth}
\begin{tabular}{c | c c c}
Bilinear $(J^\mu)$ & $SU(3)_c$ & $SU(2)_L$ &$U(1)_Y$\\
\hline
$\overline \ell_L\gamma_\mu q_L$& $ \bar 3$ & 1\,,3 &-2/3\\
$\overline e_R\gamma_\mu d_R$& $\bar 3$ & 1 &-2/3\\
$\bar\ell_L^c\gamma_\mu d_R$& $\bar 3$ & 2 &5/6\\
$\overline e_R^c\gamma_\mu q_L$& $\bar 3$ & 2 &5/6\\
$\overline e_R\gamma_\mu u_R$& $\bar 3$ & 1 &-5/3\\
$\bar\ell_L^c\gamma_\mu u_R$& $\bar 3$ & 2 &-1/6\\
\end{tabular}
\caption{Charge assignment for leptoquark vector-bosons, $V_\mu$, as a function of the SM fermion current to which they couple.\label{TabV}}
\end{minipage}
\end{figure}

All the bosons should be fundamentals of the color group and 
therefore its mass high enough not to have been produced and detected
at the LHC. As for the hyper-charges we note that the ``coincidences''
in tabs.~(\ref{TabS}) and~(\ref{TabV})
are not so but follow from the fact that Yukawa terms in the SM can be build
for quarks and leptons with the same hyper-charge $1/2$ scalar.

We shall write the Lagrangian for the bosons, respectively, as:
\begin{align}\label{Lag}
\mathcal L_\Delta&=-\Delta^\dagger\left(\left[D^2+M_{\Delta}^2\right]\Delta-y^\dagger J\right)+ J^\dagger y\, \Delta\\
\mathcal L_V&=\left(V^\mu\right)^\dagger \left(\left[D^2+M_{V}^2\right]V_\mu+g^\dagger J_\mu\right)+ \left(J^\mu\right)^\dagger\, g\, V_\mu
\end{align}
where $J$ and $J_\mu$ are the bilinears in tabs.~(\ref{TabS}) and~(\ref{TabV}),
 $D^2=D^\mu D_\mu$ is the covariant derivative containing the
SM gauge bosons, $M^2>0$ and flavor indices have been omitted for clarity. The way
the Lagrangian is written is useful for the integration of the heavy bosons;
the term in parenthesis equated to zero is the E.O.M. and vanishes on-shell. 

\subsection{Contributions to low energy processes}

Integrating out the leptoquark bosons in eq.~(\ref{Lag}) yields formally the following effective Lagrangian,
\begin{align}
\mathcal {L}_{\rm eff}&=
J^\dagger y \frac{1}{M_{\Delta}^2}y^\dagger J +\mathcal{O}\left(\frac{1}{M_{\Delta}^4}\right)\\
\mathcal {L}_{\rm eff}&=
-J^\dagger_\mu\, g \frac{1}{M_{V}^2}g^\dagger J^\mu +\mathcal{O}\left(\frac{1}{M_{V}^4}\right)
\end{align}
which can be projected in basis of operators of the SM, as is done in tab.~(\ref{tab:matching}) and, after EWSB, contributes to
the $B$-meson semi-leptonic Lagrangian as specified in tab.~(\ref{Tab:LQbsll}).

\bibliography{LQ-Bosons-Bibv7}
\end{document}